\documentclass[12pt]{article}
\usepackage{amssymb}
\usepackage{amsfonts}
\usepackage{amsmath}
\usepackage{titletoc}
\usepackage[abbr]{harvard}
\usepackage{setspace}
\usepackage{geometry}
\usepackage{float}

\newtheorem{theorem}{Theorem}

\newtheorem{condition}{Condition}

\newtheorem{lemma}{Lemma}

\newtheorem{proposition}[theorem]{Proposition}

\input{tcilatex}
\let\chapter\undefined
\renewcommand{\cite}{\citeasnoun}
\renewcommand{\thepage}{}
\renewcommand{\thefootnote}{\fnsymbol{footnote}}
\allowdisplaybreaks
\renewcommand{\baselinestretch}{1.3}

\begin{document}

\title{%
\vspace{-6ex}%
Estimation and Inference about Tail Features with Tail Censored Data\thanks{%
We thank Ulrich M\"{u}ller for very inspiring discussions. We also thank Tim
Christensen, Bo Honor\'{e}, Jia Li, Andrew Patton, Alexis Toda, Mark Watson,
and participants at numerous seminar/conference presentations for very
helpful comments.}}
\author{\textsc{Yulong Wang}\thanks{\textit{Address}: Department of
Economics, Syracuse University, 127 Eggers Hall, Syracuse, NY 13244. \textit{%
E-mail}: \texttt{ywang402@maxwell.syr.edu}} \\
Syracuse University \and \textsc{Zhijie Xiao}\thanks{\textit{Address}:
Department of Economics, Boston College, 380 Maloney Hall, Chestnut Hill, MA
02467. \textit{E-mail}: \texttt{zhijie.xiao@bc.edu}} \\
Boston College}
\date{First version: May 2019\\
This version: February 2020}
\maketitle

\begin{abstract}
This paper considers estimation and inference about tail features when the
observations beyond some threshold are censored. We first show that ignoring
such tail censoring could lead to substantial bias and size distortion, even
if the censored probability is tiny. Second, we propose a new maximum
likelihood estimator (MLE) based on the Pareto tail approximation and derive
its asymptotic properties. Third, we provide a small sample modification to
the MLE by resorting to Extreme Value theory. The MLE with this modification
delivers excellent small sample performance, as shown by Monte Carlo
simulations. We illustrate its empirical relevance by estimating (i) the
tail index and the extreme quantiles of the US individual earnings with the
Current Population Survey dataset and (ii) the tail index of the
distribution of macroeconomic disasters and the coefficient of risk aversion
using the dataset collected by \cite{Barro08}. Our new empirical findings
are substantially different from the existing literature.

\addtocontents{toc}{nprotectnenlargethispage*{1000pt}} \flushleft\textbf{%
Keywords:} Extreme Value theory, power law, extreme quantile, tail index%
\addtocontents{toc}{nprotectnpagebreak}%
\end{abstract}

\setcounter{page}{1} \renewcommand{\thepage}{\arabic{page}} %
\renewcommand{\thefootnote}{\arabic{footnote}} 
\renewcommand{\baselinestretch}{1.2} \small \normalsize%

\section{Introduction}

Tail risk and extreme events are important research topics in economics and
finance. In many applications, the features of interest are tail properties
such as tail index and extreme quantiles. Existing literature has
extensively studied the case with fully observed datasets. In comparison,
this article explores the case with tail censoring. We argue that it is
important to take into account the censoring if the research interest is in
the tail, even when the censoring fraction is small. We provide a new method
to construct estimators and confidence intervals for tail features.

Suppose one has a random sample from some underlying distribution $F$, where
the observations larger than some threshold $T$ are replaced with $T$ or
simply unobserved. In principle, tail features cannot be even identified if
they entirely depend on the right tail part of $F$ that is beyond $T$.
However, we can back out the tail-related features by extrapolation under
two assumptions. They are that (i) the tail of $F$ can be well approximated
by some suitably chosen parametric distribution, and (ii) $T$ is
sufficiently large so that only a small fraction of samples are censored.
The first assumption has been thoroughly studied in the statistic literature
and is satisfied by many commonly used distributions. The second assumption
is also satisfied in many interesting empirical applications, which motivate
this article.

Our first motivating example is the Current Population Survey (CPS) dataset,
which has been the primary data source used for investigating the
distributions of individual earnings and household income in the US.
Featured studies using CPS data include \cite{Armour13} and \cite{Eika19},
among many others. In CPS, the individual earnings larger than some
threshold $T$ are typically censored (also called topcoded) and replaced
with $T$ for confidential reasons.\footnote{%
The topcoding has constantly been changing. Description of the topcoding
mechanism is available at https://cps.ipums.org/cps/topcodes\_tables.} In
2019, the censoring threshold is 310000 USD, leading to an approximately
0.58\% censoring fraction in the full sample of individuals between 18 and
70 years old. This quantity is also substantially different across the
subsamples defined by race and gender but remains small, as seen in Table %
\ref{tbl intro}. Using this dataset, we aim to estimate and construct
confidence intervals for the tail index that measures the tail heaviness of
the income distribution and the extreme quantiles.

\begin{table}[H]
\begin{center}%
\caption{Fractions and Numbers of Censored Observations in the 2019 CPS
Dataset}\label{tbl intro}%
\vspace{+2ex}%

\begin{tabular}{ccccccc}
\hline\hline
& $n$ & {\small cen\%} & {\small cen\#} & $n$ & {\small cen\%} & {\small %
cen\#} \\ \hline
\multicolumn{1}{l}{\small full sample} & \multicolumn{1}{r}{\small 115424} & 
{\small 0.582} & {\small 672} &  &  &  \\ 
{\small race\TEXTsymbol{\backslash}gender} & \multicolumn{3}{c}{\small male}
& \multicolumn{3}{c}{\small female} \\ \hline
\multicolumn{1}{l}{\small all} & \multicolumn{1}{r}{\small 55553} & {\small %
0.884} & {\small 491} & \multicolumn{1}{r}{\small 59871} & {\small 0.302} & 
{\small 181} \\ 
\multicolumn{1}{l}{\small white} & \multicolumn{1}{r}{\small 43371} & 
{\small 0.966} & {\small 419} & \multicolumn{1}{r}{\small 45424} & {\small %
0.310} & {\small 141} \\ 
\multicolumn{1}{l}{\small Asian} & \multicolumn{1}{r}{\small 3676} & {\small %
1.360} & {\small 50} & \multicolumn{1}{r}{\small 4099} & {\small 0.537} & 
{\small 22} \\ 
\multicolumn{1}{l}{\small Hispanic} & \multicolumn{1}{r}{\small 44420} & 
{\small 1.002} & {\small 445} & \multicolumn{1}{r}{\small 48192} & {\small %
0.322} & {\small 155} \\ 
\multicolumn{1}{l}{\small black} & \multicolumn{1}{r}{\small 6144} & {\small %
0.195} & {\small 12} & \multicolumn{1}{r}{\small 7827} & {\small 0.115} & 
{\small 9} \\ \hline
\end{tabular}

\vspace{-4ex}%

\end{center}
\begin{singlespacing}%
\begin{footnotesize}%
Note: Entries are the sample sizes ($n$), the fractions in percentage
(cen\%) and the numbers (cen\#) of censored observations in individual
earnings from the March CPS variable ERN\_VAL. Data are available at
https://usa.ipums.org/usa/.%
\end{footnotesize}%
\end{singlespacing}%
\end{table}%

In our second application, we examine the size distribution of macroeconomic
disasters, which is investigated by \cite{Barro08} and \cite{Barro11}. In
particular, \cite{Barro11} define a macroeconomic disaster if the annual
Gross Domestic Product (GDP) (or consumption) declines by more than 10\%.
The authors collect the data in 36 countries from 1870 to 2005 and construct
a sample of approximately 5000 observations. However, data are missing for
four countries during WWII due to government collapse or fighting wars.
These observations correspond to the end-of-world case, and hence \cite%
{Barro11} concern that they are the largest observations but censored. In
this situation, the tail censoring fraction is about 0.1\%, and the
parameter of interest is the tail index and the coefficient of risk
aversion. Note that the censoring threshold $T$ is unknown here.

One would think that a tiny censoring fraction makes nearly no effect if we
ignore it. This is true if the object of interest lies in the mid-sample,
such as the median. However, such ignorance could lead to substantial bias
and size distortion if some tail features are of interest. To examine this,
we conduct an extensive Monte Carlo study and find that even the 0.1\% tail
censoring could lead to poor finite sample performance in some commonly used
methods, including, for example, the classical Hill (1975)'s estimator.%
\nocite{Hill75}

To accommodate the censoring, existing studies typically rely on some
parametric assumption of the whole distribution (e.g., \cite{Aban06}, \cite%
{Jenkins10}, and \cite{Burkhauser10}). Then, tail features can be expressed
as functions of the unknown parameters and estimated by the maximum
likelihood estimator (MLE). However, the parametric assumption on the whole
distribution may lead to a substantial misspecification error when the
object of interest is in the tail. This is because tail features such as
very large quantiles are typically on a large scale, and hence small
misspecification can be considerably amplified. For example, the standard
normal distribution and the Student-t distribution with 20 degrees of
freedom share almost the same shape in the mid-sample but exhibit
substantially different extreme quantiles. Such misspecification is
documented by \cite{Brzezinski2013} in a large-scale simulation study.

Instead of modeling the whole underlying distribution $F$, we focus on the 
\textit{tail} part only and approximate it with the generalized Pareto
distribution (GPD). Such a Pareto-tail approximation holds for many commonly
used distributions, including, for example, Student-t, F, Beta, and Gaussian
distributions. See Chapter 1 of \cite{deHaan07} for an overview. Given this
approximation, we first pick some tail cutoff $u$, such as some large
empirical quantile, and treat the observations larger than $u$ (but still
less than $T$) as stemming from the censored Pareto tail. Let $k$ denote the
number of these effective \textit{tail} observations. Then, we can fit them
into the classical Tobit model and conduct the MLE of the unknown
parameters. Under some mild regularity conditions, we show that the MLE is
consistent and asymptotically normal as $k$ diverges, enabling the
construction of confidence intervals. Then we can estimate extreme quantiles
by expressing them as functions of the Pareto parameters. This is formally
studied in Section \ref{sec inck}.

The proposed maximum likelihood method can be quite good in finite samples
for some applications, but it is also easy to find examples where the
asymptotic distributions provide poor approximations. A fundamental
limitation of our MLE and many other existing approaches in studying tail
features is that they require restrictive conditions for the choice of $k$
(and equivalently $u$). On the one hand, $k$ has to be sufficiently large to
support enough observations stemming from the approximately Pareto tail for
the consistency and the asymptotic Gaussianity. On the other hand, $k$ has
to be sufficiently small relative to the whole sample size $n$ so that the
tail Pareto approximation incurs a negligible bias. Such a delicate balance
is technically reflected in the conditions that $k\rightarrow \infty $ and $%
k/n\rightarrow 0$. As such, for some combinations of $n$ and $F$, it is hard
to find the $k$ that leads to satisfactory inference. This situation is
close in spirit to the bias-variance trade-off in choosing the bandwidth in
the standard kernel regressions.

To alleviate the above issue, we propose a small sample modification of the
MLE when $k$ is only moderate, say 100. In particular, we follow \cite%
{MuellerWang17} to consider $k$ as a fixed number and study the fixed-$k$
asymptotic embedding by resorting to Extreme Value (EV) theory. Instead of
treating the tail observations as \textit{independent} copies from the GPD,
EV theory treats them as \textit{dependent} random variables with a joint EV
distribution. Such dependence is negligible when $k$ is large but plays an
important role when $k$ is only moderate. Using the EV approximation, we
propose new confidence intervals for the tail index and extreme quantiles,
which have excellent coverage probabilities. This is studied in detail in
Section \ref{sec fk}.

In summary, the method that we propose is a hybrid approach. We suggest
using the MLE for estimation and inference when $k$ (and $n$) is large
enough and switching to the fixed-$k$ intervals otherwise. We choose the
switching cutoff to be $k\gtrless 250$ based on our Monte Carlo experiments
in Section \ref{sec mc}.

Returning to the CPS application, $n$ is large enough in the full sample to
support a large $k$, and hence the MLE is expected to perform well. However,
in the Asian male subsample, $n$ is only 3676. Then choosing $k$ as a small
fraction, say 5\%, of $n$, leads to only 180 tail observations and triggers
the switching. By using the new approach, we make several interesting
empirical findings. First, the tail index is substantially different across
genders and races, while the existing literature commonly focuses on the
full sample and finds the tail index to be approximately 0.5. See \cite%
{TodaWang2019} and references therein. Second, extreme quantiles also
considerably vary across genders and races. In particular, the 99.9\%
quantile of all males can be twice larger than that of the black male group.
Third, the tail features are also substantially different across ages. The
middle-aged groups exhibit heavier tails and larger extreme quantiles than
the groups below 30 or above 60 years old.

In the macroeconomic disaster application, we use the new method to
construct confidence intervals for the tail index and the coefficient of
risk aversion. We find substantially different results from those in \cite%
{Barro11}. In particular, we obtain a significantly heavier tail in the
disaster distribution. This further results in a smaller coefficient of
relative risk aversion, approximately 0.75 instead of 3. Our Monte Carlo
simulation statistically justifies such a vast difference.

The rest of the paper is organized as follows. Section \ref{sec inck}
develops the MLE, and Section \ref{sec fk} provides the small sample
modification. Section \ref{sec mc} reports Monte Carlo simulations, and
Section \ref{sec emp} applies the new approach to the CPS and the
macroeconomic disaster examples. Section \ref{sec conclusion} concludes with
some remarks. All proofs and computational details are collected in the
Appendix.

\section{The Maximum Likelihood Estimator\label{sec inck}}

Consider a random sample $\{Y_{i}\}_{i=1}^{n}$ generated from some
cumulative distribution function (CDF) $F$. Due to censoring, the
econometrician observes the pair $\left( Y_{i}^{0},D_{i}\right) ^{\intercal
} $ such that 
\begin{eqnarray}
Y_{i}^{0} &=&D_{i}T+\left( 1-D_{i}\right) Y_{i}  \label{dgp} \\
D_{i} &=&\mathbf{1}\left[ Y_{i}>T\right] ,  \notag
\end{eqnarray}%
where $T$ denotes some constant censoring threshold and $\mathbf{1}\left[
\cdot \right] $ the indicator function. Without loss of generality, we focus
on the right tail. Define $m=\dsum_{i=1}^{n}D_{i}$ as the number of censored
observations. We assume the density of $Y_{i}$, denoted as $f(\cdot )$, is
continuous and positive so that $\mathbb{P}\left( Y_{i}=T\right) =0$.

The model (\ref{dgp}) has spawned a vast literature about estimation and
inference about the mid-sample features, such as median, non-extreme
quantiles, and regression coefficients. See, for example, \cite{Powell1986}, 
\cite{Portnoy2003}, and \cite{HongTamer2003}. These mid-sample features are
typically estimated at the root-$n$ rate. In contrast, the tail features are
estimated at a much slower rate since only the largest observations are
informative about the right tail.

Define 
\begin{equation*}
F_{u}\left( y\right) =\frac{F\left( u+y\right) -F\left( u\right) }{1-F\left(
u\right) }
\end{equation*}%
as the conditional CDF given that $Y_{i}$ is larger than some pre-specified
tail cutoff $u$. We aim to approximate $F_{u}\left( y\right) $ by the
generalized Pareto distribution, which is given by%
\begin{equation}
G\left( y;\xi ,\sigma \right) =\left\{ 
\begin{array}{lcc}
1-\left( 1+\frac{\xi y}{\sigma }\right) ^{-1/\xi } &  & \xi \neq 0 \\ 
1-\exp \left( -y/\sigma \right) &  & \xi =0%
\end{array}%
\right. \text{ }  \label{GPD}
\end{equation}%
with $y\in 
\mathbb{R}
^{+\text{ }}$if $\xi \geq 0$ and $y\in \left( 0,-\sigma /\xi \right) $
otherwise. Denote $y_{0}$ as the right end-point of the support of $Y_{i}$.
It is well established in the statistic literature (e.g., \cite%
{BalkemadeHaan1974} and \cite{Pickands1975}) that the GPD is a good
approximation of $F$ in the tail, in the sense that%
\begin{equation}
\lim_{u\rightarrow y_{0}}\sup_{0<y<y_{0}-u}\left\vert F_{u}\left( y\right)
-G\left( y;\xi ,\sigma \right) \right\vert =0  \label{GPD approx}
\end{equation}%
for some scale $\sigma $ implicitly depending on $u$, if and only if $F$ is
in the domain of attraction of one of the three limit laws. The parameter $%
\xi $ is referred to as the tail index, which is uniquely determined by $F$
and characterizes its tail heaviness. See Chapter 1 of \cite{deHaan07} for
an overview.

The tail approximation (\ref{GPD approx}) is a mild assumption as it is
satisfied by many commonly used distributions. In particular, the positive $%
\xi $ case covers distributions with a Pareto-type tail such as Pareto,
Student-t, and F distributions.\footnote{%
In the standard Pareto distribution with the CDF $\mathbb{P}(Y>y)\propto
y^{-\alpha }$, the tail index $\xi $ equals $1/\alpha $. We focus on $\xi $
instead of $\alpha $ for notational ease.} The case with $\xi =0$ covers the
distributions with finite moments of any order. Leading examples are normal
and log-normal distributions. The situation with a negative $\xi $ includes
the distributions with a finite right end-point. For expositional
simplicity, we focus our discussion on the case with $\xi >0$, which covers
the empirical applications with heavy tails.

In practice, we usually choose $u$ as some large order statistic of $Y_{i}$,
say the 95\% empirical quantile. We let $T=T_{n}$ and $u=u_{n}$ depend on
the sample size $n$ and assume $T_{n}>u_{n}$ (otherwise there is no
observation). Also, denote $k$ as the number of the observations between $%
u_{n}$ and $T_{n}$ and $\{Y_{(1)}\geq Y_{(2)}\geq ,\ldots ,\geq Y_{(n)}\}$
the order statistics\footnote{%
This is different from the conventional notation for order statistics, that
is, $\{Y_{n:n}\geq Y_{n:n-1}\geq ,\ldots ,\geq Y_{n:1}\}$. We think this
alternative is more intuitive in our setup, especially in Section \ref{sec
fk}.} by descending sorting. Then effectively the available observations are
the censored largest $m+k$ order statistics 
\begin{equation}
\left( \underset{m}{\underbrace{T_{n},\ldots ,T_{n}}},Y_{(m+1)},\ldots
,Y_{(m+k)}\right) ^{\intercal },  \label{Y}
\end{equation}%
where the largest $m$ order statistics, $\{Y_{(1)},\ldots ,Y_{(m)}\}$ are
censored. Using (\ref{GPD approx}), we can write the conditional
log-likelihood of the tail observations as%
\begin{align*}
\mathcal{L}_{n}\left( \xi ,\sigma \right) & =\dsum_{i=1}^{m+k}\left\{
D_{i}\log \left( 1-F_{u}\left( T_{n}-u_{n}\right) \right) +\left(
1-D_{i}\right) \log \frac{f\left( Y_{(i)}-u_{n}\right) }{1-F\left(
u_{n}\right) }\right\} \\
& \approx \dsum_{i=1}^{m+k}\left\{ D_{i}\log \left( 1-G\left(
T_{n}-u_{n}\right) \right) +\left( 1-D_{i}\right) \log g\left(
Y_{(i)}-u_{n};\xi ,\sigma \right) \right\} \\
& =\dsum_{i=1}^{m+k}\left\{ -\frac{D_{i}}{\xi }\log \left( 1+\frac{\xi
\left( T_{n}-u_{n}\right) }{\sigma }\right) -\left( 1-D_{i}\right) \log
\sigma \right. \\
& \text{ \ \ \ \ \ \ \ \ \ \ \ \ \ \ \ \ \ }\left. -\left( 1-D_{i}\right)
\left( 1+\frac{1}{\xi }\right) \log \left( 1+\frac{\xi \left(
Y_{(i)}-u_{n}\right) }{\sigma }\right) \right\} ,
\end{align*}%
where $g\left( y;\xi ,\sigma \right) =\partial G\left( y;\xi ,\sigma \right)
/\partial y$. Then the MLE of $\xi $ and $\sigma $ are constructed as%
\begin{equation}
\left( \hat{\xi},\hat{\sigma}\right) ^{\intercal }=\arg \max_{\left(
0,\infty \right) ^{2}}\mathcal{L}_{n}\left( \xi ,\sigma \right) .
\label{MLE}
\end{equation}

To derive the asymptotic properties of the MLE, we make the following
assumptions. To simplify notations, we write $\alpha =1/\xi $ when
convenient and define $L\left( y\right) =y^{\alpha }(1-F\left( y\right) )$.

\begin{condition}
\label{cond iid}$\left( Y_{i}^{0},D_{i}\right) ^{\intercal }$ is
independently and identically generated from (\ref{dgp}).
\end{condition}

\begin{condition}
\label{cond Hall82}$F(\cdot )$ is continuously differentiable with $0\left.
<\right. f(\cdot )\left. <\right. \bar{f}$ for some constant $\bar{f}<\infty 
$ and satisfies $L\left( y\right) =C(1+\delta y^{-\beta }+o\left( y^{-\beta
}\right) )$ for some constants $\beta >0$, $C\neq 0$ and $\delta \in 
\mathbb{R}
$.
\end{condition}

\begin{condition}
\label{cond top}$T_{n}\rightarrow \infty $ and $T_{n}/u_{n}\rightarrow
\kappa \in \left( 1,\infty \right) .$
\end{condition}

\begin{condition}
\label{cond kn}$k\rightarrow \infty $ and $k=o\left( n^{2\beta /\left(
\alpha +2\beta \right) }\right) .$
\end{condition}

Condition \ref{cond iid} assumes a random sample generated with censoring.
Condition \ref{cond Hall82} is imposed by \cite{Hall82}, which states that
the Pareto tail approximation involves a second-order bias of the order $%
y^{-\beta }$. This is imposed to avoid technical complexity and can be
relaxed with other weaker conditions (cf.\ \cite{Goldie87}). Consider the
Student-t distribution with zero mean, unit variance, and $v$ degrees of
freedom, for example. The CDF\ is given by 
\begin{equation*}
1-F_{t(v)}(y)=Cy^{-v}(1+\delta y^{-2}+O(y^{-4}))\text{ as }y\rightarrow
\infty \text{.}
\end{equation*}%
Then Condition \ref{cond Hall82} holds with $\xi =1/v$ and $\beta =2$.

Condition \ref{cond top} assumes the censoring threshold is larger than the
tail cutoff. Specifically, the censoring is asymptotically negligible if $%
\kappa =\infty $, and leads to no tail observation if $\kappa =1$. Condition %
\ref{cond kn} specifies the choice of the tail cutoff and equivalently the
number of tail observations $k$. The very last assumption that $k=o\left(
n^{2\beta /\left( \alpha +2\beta \right) }\right) $ satisfies $%
k/n\rightarrow 0$ and is imposed for expositional simplicity. We can relax
it into $kn^{-2\beta /\left( \alpha +2\beta \right) }\rightarrow \mu $ for
some $\mu \in 
\mathbb{R}
$ as in \cite{Smith87} and Chapter 4 of \cite{deHaan07}. Doing so leads to a
non-zero mean in the asymptotic normal distribution, which further depends
on $\mu $, $\beta $, and $\delta $. These nuisance parameters are hardly
estimable in practice, and hence researchers typically choose a sufficiently
small $k$ to retain the asymptotic zero mean. This is similar in spirit to
the undersmoothing in standard kernel regressions.

Under these conditions, the following proposition establishes the asymptotic
normality of the MLE.

\begin{proposition}
\label{prop GPDindex}Suppose Conditions \ref{cond iid}-\ref{cond kn} hold.
Then 
\begin{equation*}
k^{1/2}\binom{\hat{\xi}-\xi }{\frac{\hat{\sigma}}{\sigma }-1}\overset{d}{%
\rightarrow }\mathcal{N}\left( 0,M^{-1}\right) ,
\end{equation*}%
where the elements of $M$ are given by 
\begin{eqnarray*}
M_{11} &=&\frac{2}{\left( 1+\xi \right) (1+2\xi )}+\frac{\kappa ^{-2-1/\xi }%
}{(1+\xi )(1+2\xi )\xi ^{2}}\times \\
&&\left\{ -1-\xi +\kappa (2+4\xi )-\kappa ^{2}(1+\xi )(1+2\xi )\right\} \\
M_{22} &=&\frac{1}{\left( 1+2\xi \right) }-\frac{\kappa ^{-2-1/\xi }}{\left(
1+2\xi \right) } \\
M_{12} &=&\frac{1}{\left( 1+\xi \right) (1+2\xi )}+\frac{\kappa ^{-2-1/\xi }%
}{\left( 1+\xi \right) (1+2\xi )\xi ^{2}}\times \\
&&\left\{ -\left( 1+\xi \right) ^{2}+\left( 1-2\kappa \right) \left( 1+\xi
\right) (1+2\xi )+\kappa \left( 2+\xi \right) \left( 1+2\xi \right) \right\}
.
\end{eqnarray*}
\end{proposition}

The tail censoring complicates the asymptotic variance substantially as
compared with the no censoring case (cf.\ \cite{Smith87}). In particular,
when the censoring is asymptotically negligible ($\kappa =\infty $), the
information matrix reduces to 
\begin{equation*}
M=\left[ 
\begin{array}{cc}
\frac{2}{\left( 1+\xi \right) (1+2\xi )} & \frac{1}{\left( 1+\xi \right)
(1+2\xi )} \\ 
\frac{1}{\left( 1+\xi \right) (1+2\xi )} & \frac{1}{\left( 1+2\xi \right) }%
\end{array}%
\right] .
\end{equation*}

Given the MLE of $\xi $ and $\sigma $, we can further estimate the extreme
quantile $Q\left( 1-p\right) \equiv \inf \{y:1-p\leq F\left( y\right) \}$.
We set $p=p_{n}\rightarrow 0$ to capture the extremeness. The estimator can
be constructed by inverting (\ref{GPD}), that is, 
\begin{equation*}
\hat{Q}\left( 1-p_{n}\right) =u_{n}+\frac{\hat{\sigma}}{\hat{\xi}}\left(
d_{n}^{\hat{\xi}}-1\right) ,
\end{equation*}%
where $d_{n}=\left( m+k\right) /\left( p_{n}n\right) $. To derive a
non-trivial asymptotic result, we let $d_{0}\equiv \lim_{n\rightarrow \infty
}d_{n}>0$ so that the target quantile is of the same or larger magnitude of $%
u_{n}$ (otherwise it can be estimated by the corresponding empirical
quantile). The following proposition derives the asymptotic distribution of $%
\hat{Q}\left( 1-p_{n}\right) $.

\begin{proposition}
\label{prop GPDquan}Suppose Conditions \ref{cond iid}-\ref{cond kn} hold. If 
$d_{0}>0$, then 
\begin{equation*}
k^{1/2}\frac{\hat{Q}\left( 1-p_{n}\right) -Q\left( 1-p_{n}\right) }{\sigma
q_{\xi }\left( d_{n}\right) }\overset{d}{\rightarrow }\mathcal{N}\left(
0,\Sigma \right)
\end{equation*}%
where $q_{\xi }\left( t\right) =\xi ^{-1}t^{\xi }\log t$ and 
\begin{equation*}
\Sigma =\left( 1,\frac{d_{0}^{\xi }-1}{\xi q_{\xi }\left( d_{0}\right) }%
\right) M^{-1}\left( 1,\frac{d_{0}^{\xi }-1}{\xi q_{\xi }\left( d_{0}\right) 
}\right) ^{\intercal }+q_{\xi }\left( d_{0}\right) ^{-2}.
\end{equation*}
\end{proposition}

Proposition \ref{prop GPDquan} establishes the asymptotic normality of the
extreme quantile estimator. Then the confidence intervals for $\xi $ and $%
Q\left( 1-p_{n}\right) $ can be constructed by plugging in the estimators
for the asymptotic variance.

\section{Small Sample Modification under the Fixed-$k$ Asymptotics\label{sec
fk}}

The results in Section 2 suggest that the asymptotic normal approximation
can be used for inference about the tail features as $k$ goes to $\infty $.
In practice, however, the choice of the tail sample size $k$ is widely
accepted as a challenging question even without censoring. This is because a
good selection of $k$ has to balance the tail approximation bias and the
variance delicately. Ultimately, the underlying distribution has to be
reasonably close to the Pareto distribution in the tail to guarantee a
satisfactory finite sample performance.

The asymptotic approximation can be quite accurate for some cases, but it is
also easy to find examples where the limiting normal distribution provides a
poor approximation. Consider the example that $F$ is a mixture of the
standard normal distribution\ with probability 0.8 and some Pareto
distribution with probability 0.2. Such a mixture structure implies that
only the very few largest observations are informative about the true\ tail.
In this case, choosing a large $k$ means including too many contaminating
observations from the mid-sample, while choosing a small $k$ invalidates the
asymptotic Gaussianity. In\ principle, there is no such a procedure that
consistently justifies whether a given $k$ is appropriate when $F$ is
entirely unknown. See Theorem 5.1 in M\"{u}ller and Wang (2017) for a
discussion on the non-censored case.

Therefore, in this article, we do not focus on the choice of $k$ but instead
treat it as given.\ In some cases, $k$ is determined by some economic theory
or empirical guidance. For example, in the macroeconomic disaster
application, the economic definition of disasters for more than 10\% of GDP
decline yields the choice of $k$. In other cases, we may employ some
data-driven algorithms that balance the Pareto approximation bias and the
variance. See, for example, \cite{Hall82}, \cite{Drees01}, and \cite%
{Clauset09}.

When $k$ and $n/k$ are both sufficiently large, we would expect the MLE in (%
\ref{MLE}) based on the increasing-$k$ asymptotics to work well.
Nevertheless, $k$ is only moderate in some situations, including our
macroeconomic disaster application and the Asian male subsample in CPS. This
causes a small sample issue that the asymptotic Gaussianity is questionable.
To find a better alternative, we resort to the asymptotic embedding that
requires $n$ diverges, but $k$ remains a fixed constant. Under this fixed-$k$
asymptotic framework, the consistent estimation of the tail index and the
extreme quantiles are out of the question since the tail sample size is
fixed. Fortunately, inference about these tail features is still
implementable, as we discuss in this section.

We first study the tail index $\xi $. EV theory (the
Fisher--Tippett--Gnedenko theorem) suggests that when the underlying
distribution is within the maximum domain of attraction (e.g., Chapter 1 of 
\cite{deHaan07}), the sample maximum is asymptotically distributed as the EV
distribution, which is parametric and entirely characterized by $\xi $.
Specifically, our Condition \ref{cond Hall82} is sufficient for the maximum
domain of attraction assumption. Then, EV theory implies that there exist
sequences of constants $a_{n}$ and $b_{n}$ such that, up to some location
and scale normalization, 
\begin{equation}
\frac{Y_{(1)}-b_{n}}{a_{n}}\overset{d}{\rightarrow }X_{1},  \label{evt_1}
\end{equation}%
where the CDF of $X_{1}$ is given by%
\begin{equation}
V_{\xi }(x)=\left\{ 
\begin{array}{lcl}
1-\exp \left( -\left( 1+\xi x/\sigma \right) ^{-1/\xi }\right) &  & \xi \neq
0 \\ 
1-\exp \left( -\exp \left( -x/\sigma \right) \right) &  & \xi =0.%
\end{array}%
\right.  \label{evt dist}
\end{equation}%
We can subsume $\sigma $ into $a_{n}$ so that $\sigma $ is always $1$ in (%
\ref{evt dist}). In additional to the sample maximum, EV theory also extends
to the first $m+k$ order statistics such that if (\ref{evt_1}) holds, then
for any fixed $m$ and $k$, 
\begin{equation}
\left( \frac{Y_{(1)}-b_{n}}{a_{n}},...,\frac{Y_{(m+k)}-b_{n}}{a_{n}}\right)
^{\intercal }\overset{d}{\rightarrow }\left( X_{1},...,X_{m+k}\right)
^{\intercal }.  \label{evt_conv}
\end{equation}%
The joint density of $\mathbf{(}X_{1},\ldots ,X_{m+k}\mathbf{)}^{\intercal }$
is given by $V_{\xi }(x_{m+k})\prod_{i=1}^{m+k}v_{\xi }(x_{i})/V_{\xi
}(x_{i})$ on $x_{m+k}\leq x_{m+k-1}\leq \ldots \leq x_{1}$, where $v_{\xi
}(x)=dV_{\xi }(x)/dx$.

Since the first $m$ elements are censored, the effective tail observations
asymptotically reduce to%
\begin{equation*}
\mathbf{X}=\left( X_{m+1},...,X_{m+k}\right) ^{\intercal },
\end{equation*}%
whose density is derived in the following proposition.

\begin{proposition}
\label{prop evt}Suppose Conditions \ref{cond iid} and \ref{cond Hall82}
hold. Then for any fixed positive integers $m$ and $k$, there exist
sequences of constants $a_{n}$ and $b_{n}$ such that 
\begin{equation*}
\frac{\left( Y_{(m+1)},\ldots ,Y_{(m+k)}\right) ^{\intercal }-b_{n}\iota _{k}%
}{a_{n}}\text{ }\overset{d}{\rightarrow }\mathbf{X,}
\end{equation*}%
where $\iota _{k}\ $denotes the $k\times 1$ vector of ones, and the joint
density of $\mathbf{X}$ is given by%
\begin{eqnarray}
f_{\mathbf{X|}\xi }\left( x_{m+1},...,x_{m+k}\right) &=&\frac{1}{m!}\left(
-\log V_{\xi }\left( x_{m+1}\right) \right) ^{m}V_{\xi }\left(
x_{m+k}\right) \tprod_{i=m+1}^{m+k}v_{\xi }\left( x_{i}\right) /V_{\xi
}\left( x_{i}\right)  \label{fx} \\
&=&\frac{1}{m!}\exp \left( 
\begin{array}{c}
-\frac{m}{\xi }\log \left( 1+\xi x_{m+1}\right) -\left( 1+\xi x_{m+k}\right)
^{-1/\xi } \\ 
-\left( 1+\frac{1}{\xi }\right) \dsum_{i=1}^{k}\log \left( 1+\xi
x_{m+i}\right)%
\end{array}%
\right) .  \notag
\end{eqnarray}
\end{proposition}

It is clear that the elements of $\mathbf{X}$ are dependent as captured by
the term $\left( -\log V_{\xi }\left( x_{m+1}\right) \right) ^{m}V_{\xi
}\left( x_{m+k}\right) $, which is negligible if $k$ is large but plays a
vital role when $k$ is only moderate. This is the fundamental difference
between the increasing-$k$ and the fixed-$k$ asymptotic embeddings, which
are respectively used by the MLE and its small sample modification. From now
on, we use bold letters to denote vectors.

If the constants $a_{n}$ and $b_{n}$ were known, the vector 
\begin{equation*}
\mathbf{Y}=\left( Y_{(m+1)},\ldots ,Y_{(m+k)}\right) ^{\intercal }
\end{equation*}%
is then approximately distributed as $\mathbf{X}$, and the limiting problem
is reduced to the small sample parametric one: constructing a confidence
interval based on one draw $\mathbf{X}$ whose density $f_{\mathbf{X}|\xi }$
is known up to $\xi $. However, $a_{n}$ and $b_{n}$ respectively correspond
to the scale $\sigma $ and the tail location $u$. Therefore, they ultimately
depend on $F$ and are challenging to estimate. Consider the standard Pareto
distribution, for example. The Pareto exponent $\alpha $ is simply $1/\xi $.
Then the fact that $a_{n}=n^{\xi }$ implies that a small estimation bias in $%
\xi $ could be amplified by the $n$-power and lead to a poor inference.

To avoid the knowledge (and estimation) of $a_{n}$ and $b_{n}$, we consider
the following self-normalized statistics: 
\begin{eqnarray}
\mathbf{Y}^{\ast } &=&\frac{\mathbf{Y}-Y_{(m+k)}\iota _{k}}{%
Y_{(m+1)}-Y_{(m+k)}}  \label{max_inv} \\
&=&\left( 1,\frac{Y_{(m+2)}-Y_{(m+k)}}{Y_{(m+1)}-Y_{(m+k)}},...,\frac{%
Y_{(m+k-1)}-Y_{(m+k)}}{Y_{(m+1)}-Y_{(m+k)}},0\right) ^{\intercal }.  \notag
\end{eqnarray}%
It is easy to establish that $\mathbf{Y}^{\mathbf{\ast }}$ is maximal
invariant with respect to the group of location and scale transformations
(cf. Chapter 6 of \cite{Lehmann05}). In words, the statistic constructed as
a function of $\mathbf{Y}^{\ast }$ remains unchanged if data are shifted and
multiplied by any non-zero constant. This invariance is also intuitive since
the tail shape should preserve no matter how data are linearly transformed.

The continuous mapping theorem and Proposition \ref{prop evt} yield that 
\begin{equation}
\mathbf{Y}^{\ast }\overset{d}{\rightarrow }\mathbf{X}^{\ast }=\left( 1,\frac{%
X_{m+2}-X_{m+k}}{X_{m+1}-X_{m+k}},...,\frac{X_{m+k-1}-X_{m+k}}{%
X_{m+1}-X_{m+k}},0\right) ^{\intercal },  \label{X_sm}
\end{equation}%
which is again invariant to location and scale transformation. By change of
variables, the density of $\mathbf{X}^{\ast }$ is given by%
\begin{equation}
f_{\mathbf{X}^{\mathbf{\ast }}|\xi }\left( \mathbf{x}^{\mathbf{\ast }%
}\right) =\frac{\Gamma \left( k+m\right) }{m!}\int_{0}^{\infty }s^{k-2}\exp
\left( -\frac{m}{\xi }\log \left( 1+\xi s\right) \right) e\left( \mathbf{x}^{%
\mathbf{\ast }},s\right) ds,  \label{fxstar}
\end{equation}%
where $e\left( \mathbf{x}^{\mathbf{\ast }},s\right) =\exp \left( -(1+1/\xi
)\sum_{i=1}^{k}\log (1+\xi x_{i}^{\mathbf{\ast }}s)\right) $ and $x_{i}^{%
\mathbf{\ast }}$ denotes the $i$th component of $\mathbf{x}^{\mathbf{\ast }}$%
. See Appendix A.1 for more details.

The density (\ref{fxstar}) can be used to conduct inference about $\xi $. In
particular, consider the hypothesis testing problem%
\begin{equation*}
H_{0}:\xi =\xi _{0}\text{ against }H_{1}:\xi \in \Xi \backslash \{\xi _{0}\},
\end{equation*}%
where $\Xi $ denotes the parameter space of $\xi $. We set $\Xi =(0,1)$ in
the Monte Carlo simulations to cover the distributions with an unbounded
support and a finite mean, which can be easily extended. Since the
alternative hypothesis is composite, we follow \cite{Andrews94} and \cite%
{EMW15} to consider the weighted average alternative%
\begin{equation*}
\int_{\Xi }f_{\mathbf{X}^{\ast }|\xi }\left( \mathbf{x}^{\ast }\right)
dW(\xi ),
\end{equation*}%
where the weighting measure $W$ reflects the importance a researcher
attaches to different alternative values of $\xi $. In practice, we set $%
W(\cdot )$ to be the CDF of the standard uniform distribution for simplicity.

Given the density of $\mathbf{X}^{\ast }$, we construct the likelihood-ratio
test as%
\begin{equation}
\varphi \left( \mathbf{x}^{\ast }\right) =\mathbf{1}\left[ \frac{\int_{\Xi
}f_{\mathbf{X}^{\ast }|\xi }\left( \mathbf{x}^{\ast }\right) dW(\xi )}{f_{%
\mathbf{X}^{\ast }|\xi _{0}}\left( \mathbf{x}^{\ast }\right) }>\mathrm{cv}%
\left( \xi _{0},k,m\right) \right] ,  \label{LRtest}
\end{equation}%
where $\mathrm{cv}\left( \xi _{0},k,m\right) $ denotes the critical value
that depends on the significance level, the null value $\xi _{0}$, the tail
sample size $k$, and the number of the censored observations $m$. The
critical value is obtained by simulation, and the test is implemented by
replacing $\mathbf{X}^{\ast }$ with $\mathbf{Y}^{\mathbf{\ast }}$ in finite
samples. The continuous mapping theorem and Proposition \ref{prop evt} yield
that $\mathbb{E}\left[ \varphi \left( \mathbf{Y}^{\mathbf{\ast }}\right) %
\right] \rightarrow \mathbb{E}\left[ \varphi \left( \mathbf{X}^{\ast
}\right) \right] $, which equals the nominal level under the null
hypothesis. The confidence interval is constructed by inverting the test.
Note that this method does not require the knowledge of the censoring
threshold $T$, which applies to cases such as the macroeconomic disaster.

Following \cite{MuellerWang17}, we can also construct the confidence
intervals of the extreme quantiles under the fixed-$k$ asymptotics. To this
end, we focus on the $Q(1-p_{n})$ quantile with $p_{n}=O(n^{-1})$, which
captures the fact that the object of interest is of the same order of
magnitude as the sample maximum. In particular, we consider $p_{n}=h/n$ for
some fixed $h>0$. Then EV theory implies that $\left( Q\left( 1-h/n\right)
-b_{n}\right) /a_{n}$ converges to the $e^{-h}$ quantile of $X_{1}$, denoted 
$q(\xi ,h)=\left( h^{-\xi }-1\right) /\xi $. Again, the research problem
would become inference about $q(\xi ,h)$ based on the $k\times 1$ vector of
observations $\mathbf{X}$ if $a_{n}$ and $b_{n}$ were known. Without loss of
generality, we construct a confidence set $S(\mathbf{Y})\subset 
\mathbb{R}
$ such that $\mathbb{P}\left( Q\left( 1-p_{n}\right) \left. \in \right. S(%
\mathbf{Y})\right) \geq 1-\mathrm{lv}$, at least as $n\rightarrow \infty $,
where $\mathrm{lv}$ denotes the significance level.

To eliminate $a_{n}$ and $b_{n}$, we use the self-normalized vector $\mathbf{%
Y}^{\ast }$ as in (\ref{fxstar}). Besides, we also impose location and scale
equivariance on our confidence interval $S$. Specifically, we impose that
for any constants $a>0$ and $b$, our interval $S$ satisfies that $S(a\mathbf{%
Y}+b)=aS(\mathbf{Y})+b$, where $aS(\mathbf{Y})+b=\{y:(y-b)/a\in S(\mathbf{Y}%
)\}$. Under this equivariance constraint, we can write%
\begin{eqnarray*}
\mathbb{P}(Q\left( 1-p_{n}\right) \left. \in \right. S(\mathbf{Y})) &=&%
\mathbb{P}\left( \frac{Q\left( 1-p_{n}\right) -b_{n}}{a_{n}}\in S\left( 
\frac{\mathbf{Y}-b_{n}\iota _{k}}{a_{n}}\right) \right) \\
&=&\mathbb{P}\left( \frac{Q\left( 1-p_{n}\right) -Y_{(m+k)}}{%
Y_{(m+1)}-Y_{(m+k)}}\in S\left( \mathbf{Y}^{\ast }\right) \right) \\
&\rightarrow &\mathbb{P}_{\xi }\left( \frac{q(\xi ,h)-X_{m+k}}{%
X_{m+1}-X_{m+k}}\in S(\mathbf{X}^{\ast })\right) ,
\end{eqnarray*}%
where the notation $\mathbb{P}_{\xi }$ (and $\mathbb{E}_{\xi }$ below)
indicates that the randomness is entirely characterized by $\xi $
asymptotically. The asymptotic problem then is the construction of a
location and scale equivariant $S$ that satisfies%
\begin{equation}
\mathbb{P}_{\xi }\left( \frac{q(\xi ,h)-X_{m+k}}{X_{m+1}-X_{m+k}}\in S(%
\mathbf{X}^{\ast })\right) \geq 1-\mathrm{lv}\text{ for all }\xi \in \Xi
\label{asy_cov}
\end{equation}%
since any $S$ that satisfies Proposition \ref{prop evt} and the equivariance
constraint also satisfies 
\begin{equation*}
\lim \inf_{n\rightarrow \infty }\mathbb{P}(Q\left( 1-p_{n}\right) \in S(%
\mathbf{Y}))\geq 1-\mathrm{lv}.
\end{equation*}%
This problem involves a single observation $\mathbf{X}\in 
\mathbb{R}
^{k}$ from a parametric distribution indexed only by the scalar parameter $%
\xi \in \Xi $.

In principle, there could still be many solutions that satisfy the
asymptotic size constraint. To obtain the optimal one, we consider the
weighted average expected length criterion%
\begin{equation}
\int \mathbb{E}_{\xi }[\func{lgth}(S(\mathbf{X}))]dW(\xi )\text{,}
\label{asy_length}
\end{equation}%
where $W$ again denotes some weighting measure on $\Xi $, and $\func{lgth}%
(A)=\int \mathbf{1}[y\in A]dy$ for any Borel set $A\subset 
\mathbb{R}
$.

To solve the program of minimizing (\ref{asy_length}) subject to (\ref%
{asy_cov}) among all equivariant set estimators $S$, we introduce 
\begin{equation*}
Y^{\ast }(\xi )=\frac{q(\xi ,h)-X_{m+k}}{X_{m+1}-X_{m+k}},
\end{equation*}%
and write $\mathbb{E}_{\xi }[\func{lgth}(S(\mathbf{X}))]\left. =\right. 
\mathbb{E}_{\xi }[(X_{m+1}-X_{m+k})\func{lgth}(S(\mathbf{X}^{\ast }))]\left.
=\right. \mathbb{E}_{\xi }[\kappa _{\xi }(\mathbf{X}^{\ast })\func{lgth}(S(%
\mathbf{X}^{\ast }))]$ with $\kappa _{\xi }(\mathbf{X}^{\ast })=\mathbb{E}%
_{\xi }[X_{m+1}-X_{m+k}|\mathbf{X}^{\ast }]$. Thus, our problem becomes 
\begin{equation}
\begin{tabular}{l}
$\min_{S(\cdot )}\int_{\Xi }\mathbb{E}_{\xi }[\kappa _{\xi }(\mathbf{X}%
^{\ast })\func{lgth}(S(\mathbf{X}^{\ast }))]dW(\xi )$ \\ 
\multicolumn{1}{r}{$\text{s.t. }\mathbb{P}_{\xi }\left( Y^{\ast }(\xi )\in S(%
\mathbf{X}^{\ast })\right) \geq 1-\mathrm{lv}\text{ for all }\xi \in \Xi .$}%
\end{tabular}
\label{program}
\end{equation}

There are two advantages to translate the asymptotic problem into (\ref%
{program}). First, (\ref{program}) does not require the knowledge of the
censoring threshold $T$ but only the number of censored observations $m$.
Second, (\ref{program}) only involves $S$ evaluated at $\mathbf{X}^{\ast }$
and hence $\mathbf{Y}^{\ast }$ in practice. This means the knowledge of $%
a_{n}$ and $b_{n}$ is asymptotically unnecessary as long as $n$ is
sufficiently larger than $m+k$. Note that any solution to (\ref{program})
also provides the form of $S$, that is, $S(\mathbf{X})=(X_{m+1}-X_{m+k})S(%
\mathbf{X}^{\ast })+X_{m+k}$. So once $S(\cdot )$ is determined, the
confidence interval can be constructed in practice by plugging in%
\begin{equation*}
(Y_{(m+1)}-Y_{(m+k)})S(\mathbf{Y}^{\ast })+Y_{(m+k)}.
\end{equation*}

To make further progress in solving (\ref{program}), we write the problem in
the following Lagrangian form:%
\begin{equation*}
\min_{S(\cdot )}\int_{\Xi }\mathbb{E}_{\xi }[\kappa _{\xi }(\mathbf{X}^{\ast
})\func{lgth}(S(\mathbf{X}^{\ast }))]dW(\xi )+\int_{\Xi }\mathbb{P}_{\xi
}\left( Y^{\ast }(\xi )\in S(\mathbf{X}^{\ast })\right) d\Lambda (\xi ),
\end{equation*}%
where the non-negative measure $\Lambda $ denotes the Lagrangian weights
that guarantee the asymptotic size constraint. By writing the expectations
above as integrals over the densities $f_{\mathbf{X}^{\ast }\mathbf{|}\xi }$
of $\mathbf{X}^{\ast }$ and $f_{Y^{\ast }(\xi ),\mathbf{X}^{\ast }|\xi }$ of 
$(Y^{\ast }(\xi ),\mathbf{X}^{\ast })$, the solution of the above problem is
given by%
\begin{equation}
S(\mathbf{x}^{\ast })=\left\{ y:\int_{\Xi }\kappa _{\xi }(\mathbf{x}^{\ast
})f_{\mathbf{X}^{\ast }|\xi }(\mathbf{x}^{\ast })dW(\xi )<\int_{\Xi
}f_{Y^{\ast }(\xi ),\mathbf{X}^{\ast }|\xi }(y,\mathbf{x}^{\ast })d\Lambda
(\xi )\right\} .  \label{S_Lambda}
\end{equation}%
The integrals can be numerically calculated by Gaussian quadrature, and then
the only remaining challenge is to find some suitable Lagrangian weights $%
\Lambda $. We solve this challenge by the numerical approach developed in 
\cite{EMW15}. The MATLAB program and the weights $\Lambda $ are available at
the author's website. Note that $\Lambda $ only needs to be computed once by
the author instead of empirical researchers. Then the most time-consuming
part in solving the program (\ref{program}) is the numerical integration,
which costs only a few seconds in a modern PC. Further details are provided
in Appendix A.1.

\section{Monte Carlo Simulations \label{sec mc}}

This section examines the finite sample performance of the proposed method
and compares it with several popular existing methods. We generate random
samples from four commonly used distributions: the generalized Pareto
distribution with $\xi =$ $0.5$ and $\sigma =1$ (GPD), the absolute value of
the Student-t distribution with 2 degrees of freedom (\TEXTsymbol{\vert}t(2)%
\TEXTsymbol{\vert}), the F distribution with parameters 4 and 4 (F(4,4)),
and the double Pareto-lognormal distribution (dPlN), that is,%
\begin{equation*}
Y=\exp \left( c_{1}+c_{2}Z_{1}+\xi Z_{2}-c_{3}Z_{3}\right) ,
\end{equation*}%
where $Z_{1}$, $Z_{2}$, $Z_{3}$ are independent and $Z_{1}\sim N(0,1)$, and $%
Z_{2},Z_{3}\sim Exp(1)$. For parameter values, we set $c_{1}=0$, $c_{2}=0.5$%
, $\xi =0.5$, and $c_{3}=1$, which are typical values for income data as
documented in \cite{Toda12}. In particular, the dPlN distribution is the
product of independent double Pareto and lognormal variables. It has been
documented to fit well to size distributions of economic variables including
income (\cite{Reed03}), city size (\cite{Giesen10}), and consumption (\cite%
{Toda17}). In all four DGP's, the true value of the tail index is 0.5.
Regarding the tail censoring, we set the censoring threshold $T$ as the 99\%
and 99.9\% quantiles of the underlying distributions, implying that the
censored probability (cen\_p) is either 1\% or 0.1\%.

We first consider some widely used estimators in empirical studies. Due to
space limitations, we only report the results of Hill (1975)'s estimator%
\nocite{Hill75} and the bias-corrected estimator proposed by \cite{Gabaix11}
(denoted GI). The confidence intervals are based on their asymptotic
normality and the plug-in estimators of their asymptotic variances. The
sample size $n$ is 1000, 2000, and 5000, and $k$ is set as $\left[ 0.05n%
\right] $ for both methods, where $[A]$ denotes the closest integer of $A$.
All results are based on 1000 simulations.

Table \ref{tbl index Hill} depicts the mean biases and the coverage
probabilities of these two methods. Several key findings can be summarized
as follows. First, both the Hill and the GI estimators suffer from severe
biases, and the confidence intervals based on them exhibit substantial
undercoverage. This holds even if the censoring probability is only 0.1\%.
Second, ignoring the upper tail censoring tends to underestimate the tail
index, which implies a misleadingly thin tail. This is seen in Section \ref%
{sec disaster} when we study the macroeconomic disasters. Finally,
unreported results show that other methods reviewed in Chapter 3 of \cite%
{deHaan07} also suffer from substantial undercoverage. Therefore, it is
crucial to take the censoring into account, even if the censoring
probability is tiny.

\begin{table}[H]
\begin{center}%
\caption{Small Sample Properties of Estimation and Inference about Tail
Index, Igorning Tail Censoring}\label{tbl index Hill}%
\vspace{+2ex}%

\begin{tabular}{lcccccccccc}
\hline
cen\_p &  & \multicolumn{4}{c}{$1\%$} &  & \multicolumn{4}{c}{$0.1\%$} \\ 
\cline{3-6}\cline{8-11}
&  & \multicolumn{2}{c}{Bias} & \multicolumn{2}{c}{Cov} &  & 
\multicolumn{2}{c}{Bias} & \multicolumn{2}{c}{Cov} \\ \hline
n=1000 &  & Hill & GI & Hill & GI &  & Hill & GI & Hill & GI \\ 
GPD &  & -0.18 & -0.24 & 0.03 & 0.00 &  & -0.04 & -0.07 & 0.88 & 0.89 \\ 
t(2) &  & -0.16 & -0.23 & 0.10 & 0.00 &  & -0.02 & -0.06 & 0.93 & 0.93 \\ 
F(4,4) &  & -0.12 & -0.20 & 0.39 & 0.05 &  & 0.03 & -0.03 & 0.98 & 0.98 \\ 
dPlN &  & -0.17 & -0.24 & 0.05 & 0.00 &  & -0.04 & -0.04 & 0.89 & 0.90 \\ 
\hline
n=2000 &  & Hill & GI & Hill & GI &  & Hill & GI & Hill & GI \\ 
GPD &  & -0.18 & -0.24 & 0.00 & 0.00 &  & -0.04 & -0.07 & 0.85 & 0.81 \\ 
t(2) &  & -0.16 & -0.23 & 0.00 & 0.00 &  & -0.02 & -0.06 & 0.93 & 0.86 \\ 
F(4,4) &  & -0.12 & -0.20 & 0.12 & 0.00 &  & 0.03 & -0.03 & 0.97 & 0.97 \\ 
dPlN &  & -0.17 & -0.24 & 0.00 & 0.00 &  & -0.04 & -0.04 & 0.88 & 0.82 \\ 
\hline
n=5000 &  & Hill & GI & Hill & GI &  & Hill & GI & Hill & GI \\ 
GPD &  & -0.18 & -0.24 & 0.00 & 0.00 &  & -0.04 & -0.08 & 0.73 & 0.45 \\ 
t(2) &  & -0.16 & -0.23 & 0.00 & 0.00 &  & -0.02 & -0.07 & 0.90 & 0.63 \\ 
F(4,4) &  & -0.12 & -0.20 & 0.00 & 0.00 &  & 0.03 & -0.03 & 0.93 & 0.95 \\ 
dPlN &  & -0.17 & -0.24 & 0.00 & 0.00 &  & -0.03 & -0.08 & 0.77 & 0.47 \\ 
\hline
\end{tabular}

\vspace{-4ex}%
\end{center}
\begin{singlespacing}%
\begin{footnotesize}%
Note: Entries are the biases and coverage probabilities (Cov) of the 95\%
confidence intervals based on Hill's estimator (Hill) and Gabaix and
Ibragimov (2010)'s estimator (GI). Data are generated from the Pareto(0.5),
the absolute value of Student-t(2), the F(4,4), and the dPIN distributions
with the censored probability (cen\_p) being 1\% or 0.01\%. The results are
based on 1000 simulations. 
\end{footnotesize}%
\end{singlespacing}%
\end{table}%

Now we implement the new method proposed in Sections \ref{sec inck} and \ref%
{sec fk}. Table \ref{tbl index} depicts the coverage and length of the 95\%
maximum likelihood confidence intervals (denoted ml) based on Proposition %
\ref{prop GPDindex} and those of the fixed-$k$ intervals (denoted fk) by
inverting (\ref{LRtest}). Several interesting findings can be made as
follows. First, the maximum likelihood confidence intervals are
substantially longer than the fixed-$k$ ones when the sample size is not
large. Besides, the coverage probability is smaller than the nominal level
when the censoring is at the 99.9\% quantile. This is because the asymptotic
normality cannot perform well when $k$ is not large. Second, in comparison,
the fixed-$k$ ones always deliver the nominal size with shorter length,
especially when the sample size is not large. Finally, when $n$ reaches 5000
(and $k$ reaches 250), the maximum likelihood intervals are comparable with
the fixed-$k$ ones. Hence a simple rule-of-thumb choice of the switching
cutoff is $k\lessgtr 250$, provided $n$ is sufficiently large.

\begin{table}[H]
\begin{center}%
\caption{Small Sample Properties of Inference about Tail Index}\label{tbl
index}%
\vspace{+2ex}%

\begin{tabular}{lcccccccccc}
\hline
cen\_p &  & \multicolumn{4}{c}{$1\%$} &  & \multicolumn{4}{c}{$0.1\%$} \\ 
\cline{3-6}\cline{8-11}
&  & \multicolumn{2}{c}{Cov} & \multicolumn{2}{c}{Lgth} &  & 
\multicolumn{2}{c}{Cov} & \multicolumn{2}{c}{Lgth} \\ \hline
n=1000 &  & ml & fk & ml & fk &  & ml & fk & ml & fk \\ 
GPD &  & 0.98 & 0.93 & 1.39 & 0.73 &  & 0.91 & 0.95 & 0.88 & 0.70 \\ 
t(2) &  & 0.98 & 0.94 & 1.40 & 0.73 &  & 0.90 & 0.95 & 0.87 & 0.70 \\ 
F(4,4) &  & 0.99 & 0.93 & 1.39 & 0.73 &  & 0.90 & 0.95 & 0.87 & 0.70 \\ 
dPlN &  & 0.99 & 0.93 & 1.30 & 0.73 &  & 0.90 & 0.94 & 0.87 & 0.70 \\ \hline
n=2000 &  & ml & fk & ml & fk &  & ml & fk & ml & fk \\ 
GPD &  & 0.96 & 0.94 & 0.99 & 0.69 &  & 0.93 & 0.94 & 0.63 & 0.58 \\ 
t(2) &  & 0.96 & 0.93 & 0.99 & 0.69 &  & 0.92 & 0.93 & 0.62 & 0.58 \\ 
F(4,4) &  & 0.96 & 0.94 & 0.99 & 0.69 &  & 0.93 & 0.94 & 0.63 & 0.58 \\ 
dPlN &  & 0.96 & 0.94 & 0.99 & 0.69 &  & 0.92 & 0.93 & 0.62 & 0.58 \\ \hline
n=5000 &  & ml & fk & ml & fk &  & ml & fk & ml & fk \\ 
GPD &  & 0.97 & 0.94 & 0.63 & 0.54 &  & 0.95 & 0.94 & 0.40 & 0.39 \\ 
t(2) &  & 0.96 & 0.94 & 0.62 & 0.54 &  & 0.93 & 0.92 & 0.40 & 0.38 \\ 
F(4,4) &  & 0.97 & 0.95 & 0.63 & 0.54 &  & 0.94 & 0.93 & 0.40 & 0.38 \\ 
dPlN &  & 0.96 & 0.93 & 0.62 & 0.54 &  & 0.94 & 0.94 & 0.40 & 0.38 \\ \hline
\end{tabular}

\vspace{-4ex}%
\end{center}
\begin{singlespacing}%
\begin{footnotesize}%
Note: Entries are the coverage probabilities (Cov) and the averaged length
(Lgth) of the maximum likelihood intervals (ml) and the fixed-$k$ intervals
(fk) for the tail index. Data are generated from the Pareto(0.5), the
absolute value of Student-t(2), the F(4,4), and the dPIN distributions with
the censored probability (cen\_p) being 1\% or 0.01\%. The results are based
on 1000 simulations. The level of significance is 5\%. 
\end{footnotesize}%
\end{singlespacing}%
\end{table}%

Tables \ref{tbl quan99} depicts the coverage probabilities and lengths of
the confidence intervals of the 99\% quantile, using either the maximum
likelihood method as in Proposition \ref{prop GPDquan}\ or the fixed-$k$
method (\ref{S_Lambda}). Both methods deliver satisfactory size and length
properties, although the maximum likelihood intervals suffer from slight
undercoverage. However, as we target the more extreme 99.9\% quantile as in
Table \ref{tbl quan999}, such undercoverage is substantial when $k$ is less
than 250. In contrast, the fixed-$k$ ones always perform excellently. These
results reinforce our switching cutoff at $k=250$.

\begin{table}[H]
\begin{center}%
\caption{Small Sample Properties of Inference about the $0.99$ Quantile}%
\label{tbl quan99}%
\vspace{+2ex}%

\begin{tabular}{lcccccccccc}
\hline
cen\_p &  & \multicolumn{4}{c}{$1\%$} &  & \multicolumn{4}{c}{$0.1\%$} \\ 
\cline{3-6}\cline{8-11}
&  & \multicolumn{2}{c}{Cov} & \multicolumn{2}{c}{Lgth} &  & 
\multicolumn{2}{c}{Cov} & \multicolumn{2}{c}{Lgth} \\ \hline
n=1000 &  & ml & fk & ml & fk &  & ml & fk & ml & fk \\ 
GPD &  & 0.95 & 0.94 & 6.71 & 7.09 &  & 0.91 & 0.96 & 4.83 & 5.79 \\ 
t(2) &  & 0.95 & 0.94 & 6.73 & 7.13 &  & 0.91 & 0.94 & 4.87 & 5.89 \\ 
F(4,4) &  & 0.94 & 0.94 & 11.67 & 12.17 &  & 0.91 & 0.95 & 8.32 & 9.91 \\ 
dPlN &  & 0.95 & 0.94 & 4.92 & 5.25 &  & 0.91 & 0.95 & 3.54 & 4.37 \\ \hline
n=2000 &  & ml & fk & ml & fk &  & ml & fk & ml & fk \\ 
GPD &  & 0.96 & 0.94 & 4.60 & 4.86 &  & 0.92 & 0.96 & 3.38 & 3.88 \\ 
t(2) &  & 0.96 & 0.94 & 4.60 & 4.76 &  & 0.93 & 0.96 & 3.43 & 3.89 \\ 
F(4,4) &  & 0.96 & 0.95 & 8.04 & 8.09 &  & 0.92 & 0.95 & 5.85 & 6.72 \\ 
dPlN &  & 0.96 & 0.94 & 3.44 & 3.50 &  & 0.93 & 0.96 & 2.52 & 2.90 \\ \hline
n=5000 &  & ml & fk & ml & fk &  & ml & fk & ml & fk \\ 
GPD &  & 0.97 & 0.96 & 2.85 & 2.93 &  & 0.93 & 0.94 & 2.12 & 2.38 \\ 
t(2) &  & 0.97 & 0.95 & 2.84 & 2.91 &  & 0.94 & 0.96 & 2.15 & 2.40 \\ 
F(4,4) &  & 0.97 & 0.95 & 4.93 & 5.05 &  & 0.93 & 0.94 & 3.68 & 4.10 \\ 
dPlN &  & 0.97 & 0.95 & 2.08 & 2.12 &  & 0.93 & 0.96 & 1.57 & 1.77 \\ \hline
\end{tabular}

\vspace{-4ex}%
\end{center}
\begin{singlespacing}%
\begin{footnotesize}%
Note: Entries are the coverage probabilities (Cov) and the averaged length
(Lgth) of the maximum likelihood intervals (ml) and the fixed-$k$ intervals
(fk) for the 99\% quantiles. Data are generated from the Pareto(0.5), the
absolute value of Student-t(2), the F(4,4), and the dPIN distributions with
the censored probability (cen\_p) being 1\% or 0.01\%. The results are based
on 1000 simulations. The level of significance is 5\%.%
\end{footnotesize}%
\end{singlespacing}%
\end{table}%

\begin{table}[H]
\begin{center}%
\caption{Small Sample Properties of Inference about the $0.999$ Quantile}%
\label{tbl quan999}%
\vspace{+2ex}%

\begin{tabular}{lcccccccccc}
\hline
cen\_p &  & \multicolumn{4}{c}{$1\%$} &  & \multicolumn{4}{c}{$0.1\%$} \\ 
\cline{3-6}\cline{8-11}
&  & \multicolumn{2}{c}{Cov} & \multicolumn{2}{c}{Lgth} &  & 
\multicolumn{2}{c}{Cov} & \multicolumn{2}{c}{Lgth} \\ \hline
n=1000 &  & ml & fk & ml & fk &  & ml & fk & ml & fk \\ 
GPD &  & 0.86 & 0.92 & 128.9 & 102.6 &  & 0.83 & 0.95 & 61.68 & 75.08 \\ 
t(2) &  & 0.85 & 0.93 & 123.0 & 103.1 &  & 0.83 & 0.94 & 60.87 & 72.62 \\ 
F(4,4) &  & 0.85 & 0.91 & 226.8 & 178.1 &  & 0.83 & 0.95 & 106.3 & 130.8 \\ 
dPlN &  & 0.85 & 0.93 & 93.16 & 78.21 &  & 0.82 & 0.95 & 45.16 & 55.13 \\ 
\hline
n=2000 &  & ml & fk & ml & fk &  & ml & fk & ml & fk \\ 
GPD &  & 0.89 & 0.92 & 79.20 & 72.25 &  & 0.87 & 0.94 & 39.71 & 44.91 \\ 
t(2) &  & 0.88 & 0.94 & 75.60 & 71.65 &  & 0.88 & 0.95 & 39.32 & 45.96 \\ 
F(4,4) &  & 0.89 & 0.92 & 136.6 & 126.5 &  & 0.88 & 0.93 & 69.35 & 80.28 \\ 
dPlN &  & 0.88 & 0.92 & 56.17 & 51.48 &  & 0.87 & 0.93 & 28.99 & 32.38 \\ 
\hline
n=5000 &  & ml & fk & ml & fk &  & ml & fk & ml & fk \\ 
GPD &  & 0.94 & 0.95 & 43.06 & 40.41 &  & 0.92 & 0.94 & 23.96 & 24.15 \\ 
t(2) &  & 0.92 & 0.95 & 40.75 & 39.25 &  & 0.92 & 0.93 & 23.39 & 24.15 \\ 
F(4,4) &  & 0.92 & 0.95 & 73.55 & 70.90 &  & 0.91 & 0.94 & 41.11 & 41.62 \\ 
dPlN &  & 0.92 & 0.92 & 30.65 & 28.32 &  & 0.91 & 0.93 & 17.31 & 18.13 \\ 
\hline
\end{tabular}

\vspace{-4ex}%
%
\end{center}

\begin{singlespacing}%
\begin{footnotesize}%
Note: Entries are the coverage probabilities (Cov) and the averaged length
(Lgth) of the maximum likelihood intervals (ml) and the fixed-k intervals
(fk) for the 99.9\% quantiles. Data are generated from the Pareto(0.5), the
absolute value of Student-t(2), the F(4,4), and the dPIN distributions with
the censored probability (cen\_p) being 1\% or 0.01\%. The results are based
on 1000 simulations. The level of significance is 5\%.%
\end{footnotesize}%
\end{singlespacing}%
\end{table}%

\section{Empirical Applications\label{sec emp}}

Many applications in economics and finance involve estimation and inference
of tail features with censored data. In this section, we apply the proposed
method to the two datasets we discussed earlier in this paper. Our empirical
analysis highlights the potential of our approach.

\subsection{US Individual Earnings\label{sec income}}

Our first application is about the tail features of the individual earnings
distribution. Following the convention, we use the variable ERN\_VAL in the
March CPS dataset and drop the individuals that are younger than 18 or older
than 70 years old. This yields 115,424 observations in the 2019 sample. The
censoring threshold is 310000 USD, which leads to a 0.58\% censoring
fraction in the full sample and various censoring fractions in different
subsamples. The first several columns in Table \ref{tbl cps} present the
sample sizes ($n$) and the numbers (cen\#) and the fractions (cen\%) of the
censored observations, respectively. We use the previously introduced method
to construct the 95\% confidence intervals of the tail index and the 99\%
and 99.9\% quantiles. Specifically, we follow the simulation study to use
the maximum likelihood confidence intervals developed in Section \ref{sec
inck} when $k$ is larger than 250 and switch to the fixed-$k$ confidence
intervals (\ref{S_Lambda}) otherwise. The last six columns in Table \ref{tbl
cps} present the results with $k=\left[ 0.05n\right] $. The results based on
other choices are similar and reported in Appendix A.3.

Several interesting findings can be summarized as follows. First, in Panel
A, the tail index is around 0.5 in the full sample, as commonly found in the
existing literature. But it is substantially different across subsamples.
Second, the tail also exhibits substantial heterogeneity across genders. In
particular, the male sample has significantly higher quantiles than the
female at both the 99\% and 99.9\% levels. Third, this difference also
exists across races. In particular, the 99.9\% quantile of all males is at
least twice larger than that of the black males. All such heterogeneity
provides new evidence for potential racial and gender discrimination.
Finally, Panel B depicts the heterogeneity across ages, with substantially
heavier tails showing up in the middle-aged groups.

\begin{table}[H]
\begin{center}%
\caption{Empirical Results in 2019 March CPS Data}\label{tbl cps}

\begin{tabular}{lccccccccc}
\hline\hline
\multicolumn{10}{c}{Panel A: 95\% confidence intervals in
race-and-gender-based subsamples} \\ 
race-gender & $n$ & cen\# & cen\% & \multicolumn{2}{c}{tail index} & 
\multicolumn{2}{c}{Q(0.99)} & \multicolumn{2}{c}{Q(0.999)} \\ \hline
full sample & \multicolumn{1}{r}{115424} & 672 & 0.58 & (0.41 & 0.52) & 
(24.24 & 25.32) & (62.63 & 75.61) \\ 
all males & \multicolumn{1}{r}{55553} & 491 & 0.88 & (0.35 & 0.53) & (28.64
& 30.68) & (67.71 & 92.58) \\ 
all females & \multicolumn{1}{r}{59871} & 181 & 0.30 & (0.42 & 0.55) & (18.02
& 19.03) & (46.34 & 58.01) \\ 
white males & \multicolumn{1}{r}{43371} & 419 & 0.97 & (0.88 & 1.00) & (29.83
& 33.56) & (145.2 & 279.0) \\ 
white females & \multicolumn{1}{r}{45424} & 141 & 0.31 & (0.42 & 0.57) & 
(18.09 & 19.28) & (46.56 & 60.66) \\ 
Asian males & \multicolumn{1}{r}{3676} & 50 & 1.36 & (0.00 & 0.45) & (30.73
& 37.20) & (48.03 & 86.76) \\ 
Asian females & \multicolumn{1}{r}{4099} & 22 & 0.54 & (0.35 & 0.94) & (20.77
& 27.16) & (45.12 & 145.0) \\ 
Hispanic males & \multicolumn{1}{r}{44420} & 445 & 1.00 & (0.71 & 0.95) & 
(31.10 & 34.75) & (119.3 & 214.0) \\ 
Hispanic females & \multicolumn{1}{r}{48192} & 155 & 0.32 & (0.47 & 0.62) & 
(18.70 & 19.90) & (50.17 & 66.13) \\ 
black males & \multicolumn{1}{r}{6144} & 12 & 0.20 & (0.22 & 0.58) & (15.92
& 18.22) & (28.73 & 49.39) \\ 
black females & \multicolumn{1}{r}{7827} & 9 & 0.16 & (0.16 & 0.44) & (13.90
& 15.64) & (25.25 & 37.18) \\ \hline
\multicolumn{10}{c}{Panel B: 95\% confidence intervals in age-based
subsamples} \\ 
age & $n$ & cen\# & cen\% & \multicolumn{2}{c}{tail index} & 
\multicolumn{2}{c}{Q(0.99)} & \multicolumn{2}{c}{Q(0.999)} \\ \hline
18-30 & 27829 & 35 & 0.13 & (0.33 & 0.49) & (12.69 & 13.60) & (28.16 & 36.16)
\\ 
30-40 & 25213 & 158 & 0.63 & (0.28 & 0.50) & (24.12 & 26.36) & (52.24 & 
75.73) \\ 
40-50 & 23419 & 213 & 0.91 & (0.83 & 1.00) & (28.89 & 33.82) & (119.7 & 
297.2) \\ 
50-60 & 21767 & 196 & 0.90 & (0.52 & 0.83) & (28.19 & 32.21) & (78.05 & 
154.2) \\ 
60-65 & 17196 & 70 & 0.41 & (0.17 & 0.41) & (20.95 & 23.13) & (41.86 & 59.31)
\\ \hline
\end{tabular}

\vspace{-4ex}%

\end{center}
\begin{singlespacing}%
\begin{footnotesize}%
Note: Entries are the sample size ($n$), the number of censored observations
(cen\#), the censored fraction in percentage points (cen\%),\ 95\%
confidence intervals of the tail index and those of the 99\% and 99.9\%
quantiles measured in 10$^{4}$\ USD. The results are based on the variable
ERN\_VAL in the CPS dataset and equivalently the variable inclongj from the
IPUMS dataset. Data are available at https://cps.ipums.org/cps. 
\end{footnotesize}%
\end{singlespacing}%
\end{table}%

\subsection{Macroeconomic Disasters\label{sec disaster}}

This section studies the size distribution of macroeconomic disasters, which
is an important research topic in macroeconomics. \cite{Barro08} and \cite%
{Barro11} construct and analyze the dataset that consists of annual GDP (and
consumption) growth rates in 36 countries from 1870 to 2005. The authors
sort these observations and define a macroeconomic disaster if the GDP
declines by more than 10\%. This leads to $k=157$ tail observations. Then
the authors fit these data to the (double) Pareto distribution to estimate
the Pareto exponent, which is the reciprocal of the\ tail index, and back
out the coefficient of the relative risk aversion by a theoretical model
(eq.2 in \cite{Barro11}).

However, the largest disasters tend to be missing because some governments
collapsed or were fighting wars (p.1581 in \cite{Barro11}). Ignoring these
missing data in the upper tail could lead to substantial bias, as we show in
the\ Monte Carlo simulations. We revisit this problem by applying our fixed-$%
k$ method since $k$ is only moderate. Specifically, the most recent data
missing happens in four countries, which are Greece, Malaysia, the
Philippines, and Singapore during WWII. Therefore, we set $m=4$ and apply
the fixed-$k$ method to construct the 95\% confidence intervals for the tail
index $\xi $ and those for the coefficient of relative risk aversion by
solving eq.2 in \cite{Barro11}. For comparison, we also construct the
intervals based on Hill (1975)'s estimator and the bias-reduced estimator
(GI) proposed by \cite{Gabaix11}. Table \ref{empirical_index} presents the
result.

As shown in the table, the fixed-$k$ intervals contain substantially larger
values of the tail index than the other two methods that ignore the tail
censoring. This is coherent with our simulation results in Table \ref{tbl
index Hill}. By taking the reciprocal, the Pareto exponent is estimated to
be approximately 7 in \cite{Barro11} but less than 1 by the new method.
Therefore, taking the tail censoring into account leads to a substantially
heavier tail in the disaster size. Accordingly, the coefficient of risk
aversion is found to be around 0.75, which is significantly lower than 3 in 
\cite{Barro11}. These results undermine their conclusion that "the (Hill)
estimate of the upper-tail exponent is likely to have only a small upward
bias due to missing extreme observations, which have to be few in number."

\begin{table}[H]
\begin{center}%
\caption{Empirical Results in Macroeconomic Disasters}\label{empirical_index}%
\vspace{+2ex}%

\begin{tabular}{lcccccc}
\hline\hline
Method & \multicolumn{2}{c}{Hill} & \multicolumn{2}{c}{GI} & 
\multicolumn{2}{c}{New} \\ \hline
& \multicolumn{6}{c}{Tail Index} \\ 
95\% CIs & (0.12 & 0.17) & (0.16 & 0.26) & (0.57 & 1.00) \\ \hline
& \multicolumn{6}{c}{Coefficient of Risk Aversion} \\ 
95\% CIs & (3.73 & 5.10) & (2.44 & 3.88) & (0.58 & 1.04) \\ \hline\hline
\end{tabular}

\vspace{-4ex}%
\end{center}
\begin{singlespacing}
\begin{footnotesize}%
Note: Entries are 95\% confidence intervals (CIs) of the tail index of the
disaster size distribution and the coefficient of risk aversion, based on
the Hill estimator (Hill), the bias-reduced estimator proposed by \cite%
{Gabaix11} (GI), and the fixed-$k$ method by inverting (\ref{LRtest}). Data
are available at https://scholar.harvard.edu/barro/data\_sets.%
\end{footnotesize}
\end{singlespacing}%
\end{table}%

\section{Concluding Remarks\label{sec conclusion}}

This paper develops a new approach to estimate and conduct inference about
tail features for censored data. The method can be viewed as a hybrid
approach that uses the maximum likelihood estimation when the tail sample
size is large and switches to a small sample modification otherwise. As
shown in Monte Carlo simulations, the new method has excellent small sample
performance.

This new approach is empirically relevant in broad areas studying tail
features (e.g., tail index and extreme quantiles). We illustrate this with
the March CPS data and the macroeconomic disaster data and find considerably
different results from the existing literature.

There are theoretical extensions and empirical applications of our method,
which we suppress in the current paper due to space limitations. We list a
few here. First, our method naturally applies to the no censoring case by
setting $\kappa =\infty $ in the MLE and $m=0$ in the fixed-$k$ method.
Besides, we can follow \cite{MuellerWang19} to construct the (quantile)
unbiased estimation of the tail features, which could perform better in
terms of mean absolute deviation and mean squared error, especially when $k$
is not large.

Second, many other tail features can be learned by our new method as long as
they can be expressed as functions of the tail index. For example, the
conditional tail expectation is another important risk measure in finance,
which is defined as the expectation conditional on being larger than some
high quantile, that is, $\mathbb{E}\left[ Y_{i}|Y_{i}>Q(1-p)\right] $. By
reparametrizing $p=h/n$ for some $h>0$ and using EV theory, we can obtain
that that $(\mathbb{E}\left[ Y_{i}|Y_{i}>Q(1-h/n)\right] -b_{n})/a_{n}\left.
\rightarrow \right. h^{-\xi }/(\xi (1-\xi ))-1/\xi $, which again entirely
depends on $\xi $ and $h$ (p.1336 in \cite{MuellerWang19}). Then we can
construct the fixed-$k$ intervals for this quantity in an analogous fashion
to (\ref{program}).

Finally, our method also allows from weak dependent data if some additional
regularity condition is satisfied. In particular, EV theory holds under weak
dependence, such as $\alpha $-mixing, as long as the largest order
statistics do not show up in a cluster. This is referred to as the
non-cluster condition. See, for example, \cite{Leadbetter83}, \cite{Obrien87}%
, \cite{Mikosch00}, \cite{Chernozhukov05}, and \cite{Chernozhukov11}.

\section*{Appendix}

\subsection*{A.1 Computational Details}

The estimators defined in Section \ref{sec fk} require evaluation of some
expectations. Define $\Gamma \left( \cdot \right) $ as the Gamma function
and $\Gamma \left( a,x\right) =\int_{x}^{\infty }t^{a-1}e^{-t}dt$ as the
incomplete Gamma function. Also define $e\left( \mathbf{x}^{\mathbf{\ast }%
},s\right) =\exp \left( -(1+1/\xi )\sum_{i=1}^{k}\log (1+\xi x_{i}^{\mathbf{%
\ast }}s)\right) $. Change of variables and integration by parts yield that%
\begin{eqnarray*}
&&\mathbb{E}_{\xi }\left[ X_{m+1}-X_{m+k}|\mathbf{X}^{\ast }=\mathbf{x}%
^{\ast }\right] f_{\mathbf{X}^{\ast }|\xi }\left( \mathbf{x}^{\ast }\right)
\\
&=&\frac{\Gamma \left( k+m-\xi \right) }{m!}\int_{0}^{+\infty }s^{k-1}\exp
\left( -\frac{m}{\xi }\log \left( 1+\xi s\right) -(1+\frac{1}{\xi }%
)\sum_{i=1}^{k}\log (1+\xi x_{i}^{\ast }s)\right) ds.
\end{eqnarray*}%
and%
\begin{eqnarray*}
&&f_{Y^{\ast }(\xi ),\mathbf{X}^{\ast }|\xi }(y,\mathbf{x}^{\ast }) \\
&=&\frac{1}{m!}\int_{0}^{+\infty }\mathbf{1}\left[ s+\frac{x_{i}^{\ast }}{y}%
(q(\xi )-s)>0\text{ for all }i\text{ and }x_{i}^{\ast }>x_{j}^{\ast }\text{
if }i<j\right] \left\vert \left( \frac{q(\xi ,h)-s}{y}\right) ^{k-1}\frac{1}{%
y}\right\vert \\
&&\times \exp \left( 
\begin{array}{c}
-\frac{m}{\xi }\log \left( 1+\xi x+\frac{1}{y}\xi (q(\xi ,h)-x)\right)
-(1+\xi s)^{-1/\xi } \\ 
-(1+\frac{1}{\xi })\sum_{i=1}^{k}\log \left( 1+\xi s+\frac{x_{i}^{\ast }}{y}%
\xi (q(\xi )-s)\right)%
\end{array}%
\right) ds,
\end{eqnarray*}%
where $q(\xi ,h)=\left( h^{-\xi }-1\right) /\xi $. We evaluate these by
numerical quadrature.

To determine the Lagrange multipliers $\Lambda $, we use the algorithm
developed by \cite{EMW15}. In particular, we restrict $\Lambda (\cdot )$ to
be point masses with the support on the discretized $\Xi $, that is, $\Xi
_{D}=\{1/50,2/50,...,1\}$ and determine the 50 point masses by fixed-point
iterations based on Monte Carlo estimates of the coverage probabilities. To
do this, we simulate the coverage probabilities with 20000 i.i.d.\ draws
from a proposal with $\xi $ uniformly drawn from $\Xi _{D}$, and iteratively
increase or decrease the point masses on $\Xi _{D}$ as a function of whether
the coverage given that value of $\xi $ is larger or smaller than the
nominal level 0.05. Stop this iteration until the differences between the
coverages for all values of $\xi $ and 0.05 are lower than a pre-specified
tolerance $\varepsilon =0.001$. This tolerance can be arbitrarily small at
the cost of longer computation time and larger numbers of simulation draws.

For any given $k$, $h$, and $m$, the Lagrange multipliers only need to be
determined once. The tables of the Lagrange multipliers and the
corresponding MATLAB code are provided on our website:
https://sites.google.com/site/yulongwanghome/.

\subsection*{A.2 Proof}

To prove Proposition \ref{prop GPDindex}, we first establish two
intermediate results, Lemmas \ref{lemma GPD} and \ref{lemma SR2}. Throughout
the proof, we suppress the subscript $n$ in $T_{n}$ and $u_{n}$ for
notational simplicity.

\begin{lemma}
\label{lemma GPD}Suppose the conditional CDF $F_{u}$ is exactly the GPD (\ref%
{GPD}). Denote the log-likelihood as 
\begin{equation*}
l_{i}\left( \xi ,\sigma \right) =D_{i}\log \left( 1-G\left( T_{u};\xi
,\sigma \right) \right) +\left( 1-D_{i}\right) \log g\left( Y_{i};\xi
,\sigma \right) ,
\end{equation*}%
where $T_{u}=T-u$. Then the elements of the Fisher-information matrix are
given by%
\begin{eqnarray*}
M_{11} &\equiv &\mathbb{E}_{GPD}\left[ -\frac{\partial ^{2}l_{i}\left( \xi
,\sigma \right) }{\partial \xi ^{2}}\right] =\frac{2}{\left( 1+\xi \right)
(1+2\xi )}+\frac{z^{-2-1/\xi }}{(1+\xi )(1+2\xi )\xi ^{2}}\times \\
&&\left\{ -1-\xi +z(2+4\xi )-z^{2}(1+\xi )(1+2\xi )\right\} \\
M_{22} &\equiv &\mathbb{E}_{GPD}\left[ -\frac{\partial ^{2}l_{i}\left( \xi
,\sigma \right) }{\partial \sigma ^{2}}\right] =\frac{1}{\left( 1+2\xi
\right) \sigma ^{2}}-\frac{z^{-2-1/\xi }}{\left( 1+2\xi \right) \sigma ^{2}}
\\
M_{12} &\equiv &\mathbb{E}_{GPD}\left[ -\frac{\partial ^{2}l_{i}\left( \xi
,\sigma \right) }{\partial \sigma \partial \xi }\right] =\frac{1}{\left(
1+\xi \right) (1+2\xi )\sigma }+\frac{z^{-2-1/\xi }}{\left( 1+\xi \right)
(1+2\xi )\xi ^{2}\sigma }\times \\
&&\left\{ -\left( 1+\xi \right) ^{2}+\left( 1-2z\right) \left( 1+\xi \right)
(1+2\xi )+z\left( 2+\xi \right) \left( 1+2\xi \right) \right\}
\end{eqnarray*}%
where $z=1+\xi T_{u}/\sigma $.
\end{lemma}

The notation $\mathbb{E}_{GPD}$ indicates that the expectation is taken with
respect to the exact GPD in this lemma only.

\paragraph{Proof of Lemma \protect\ref{lemma GPD}}

Denote%
\begin{eqnarray}
-l_{i}\left( \xi ,\sigma \right) &=&-D_{i}\log \left( 1-G\left( T_{u};\xi
,\sigma \right) \right) -\left( 1-D_{i}\right) \log g\left( Y_{i};\xi
,\sigma \right)  \label{loglikeli} \\
&=&\frac{D_{i}}{\xi }\log \left( 1+\frac{\xi T_{u}}{\sigma }\right) +\left(
1-D_{i}\right) \left( 1+\frac{1}{\xi }\right) \log \left( 1+\frac{\xi Y_{i}}{%
\sigma }\right) +\left( 1-D_{i}\right) \log \sigma .  \notag
\end{eqnarray}%
Then by substituting $z=1+\xi T_{u}/\sigma $ and some elementary
calculation, we have 
\begin{eqnarray*}
-\frac{\partial l_{i}\left( \xi ,\sigma \right) }{\partial \xi }
&=&D_{i}\left\{ -\frac{1}{\xi ^{2}}\log z+\frac{1}{\xi ^{2}}\left(
1-z^{-1}\right) \right\} \\
&&+\left( 1-D_{i}\right) \left\{ -\frac{1}{\xi ^{2}}\log \left( 1+\frac{\xi 
}{\sigma }Y_{i}\right) +\frac{1}{\xi }\left( 1+\frac{1}{\xi }\right) \left(
1-\left( 1+\frac{\xi }{\sigma }Y_{i}\right) ^{-1}\right) \right\} , \\
-\frac{\partial l_{i}\left( \xi ,\sigma \right) }{\partial \sigma }
&=&-D_{i}\left\{ \frac{1}{\xi \sigma }\left( 1-z^{-1}\right) \right\} \\
&&+\left( 1-D_{i}\right) \left\{ -\frac{1}{\xi \sigma }+\frac{1}{\sigma }%
\left( 1+\frac{1}{\xi }\right) \left( 1+\frac{\xi }{\sigma }Y_{i}\right)
^{-1}\right\} ,
\end{eqnarray*}%
\begin{eqnarray*}
&&-\frac{\partial ^{2}l_{i}\left( \xi ,\sigma \right) }{\partial \xi ^{2}} \\
&=&D_{i}\left\{ \frac{2}{\xi ^{3}}\log z-\frac{T_{u}^{2}}{\sigma ^{2}\xi }%
z^{-2}-\frac{2T_{u}}{\sigma \xi ^{2}}z^{-1}\right\} \\
&&+\left( 1-D_{i}\right) \left\{ \frac{2}{\xi ^{3}}\log \left( 1+\frac{\xi
Y_{i}}{\sigma }\right) -\frac{3+\xi }{\xi ^{3}}+\frac{2\left( 2+\xi \right) 
}{\xi ^{3}}\left( 1+\frac{\xi }{\sigma }Y_{i}\right) ^{-1}-\frac{1+\xi }{\xi
^{3}}\left( 1+\frac{\xi Y_{i}}{\sigma }\right) ^{-2}\right\} ,
\end{eqnarray*}%
\begin{eqnarray*}
-\frac{\partial ^{2}l_{i}\left( \xi ,\sigma \right) }{\partial \sigma ^{2}}
&=&D_{i}\left\{ -\frac{T_{u}^{2}\xi }{\sigma ^{4}}z^{-2}+\frac{2T_{u}}{%
\sigma ^{3}}z^{-1}\right\} \\
&&+(1-D_{i})\left\{ \frac{1}{\xi \sigma ^{2}}-\frac{1}{\sigma ^{2}}\left( 1+%
\frac{1}{\xi }\right) \left( 1+\frac{\xi Y_{i}}{\sigma }\right) ^{-2}\right\}
\\
&=&D_{i}\left\{ \frac{1}{\sigma ^{2}\xi }\left( 1-z^{-2}\right) \right\}
+(1-D_{i})\left\{ \frac{1}{\xi \sigma ^{2}}-\frac{1}{\sigma ^{2}}\left( 1+%
\frac{1}{\xi }\right) \left( 1+\frac{\xi Y_{i}}{\sigma }\right)
^{-2}\right\} ,
\end{eqnarray*}%
and%
\begin{eqnarray*}
&&-\frac{\partial ^{2}l_{i}\left( \xi ,\sigma \right) }{\partial \sigma
\partial \xi } \\
&=&D_{i}\left\{ \frac{T_{u}^{2}}{\sigma ^{3}}\left( 1+\frac{T_{u}\xi }{%
\sigma }\right) ^{-2}\right\} +\left( 1-D_{i}\right) \left\{ \frac{1}{\sigma
\xi ^{2}}-\frac{2+\xi }{\sigma \xi ^{2}}\left( 1+\frac{Y_{i}\xi }{\sigma }%
\right) ^{-1}+\frac{\left( 1+\xi \right) }{\sigma \xi ^{2}}\left( 1+\frac{%
Y_{i}\xi }{\sigma }\right) ^{-2}\right\} \\
&=&D_{i}\left\{ \frac{1}{\sigma \xi ^{2}}\left( 1-2z^{-1}+z^{-2}\right)
\right\} +\left( 1-D_{i}\right) \left\{ \frac{1}{\sigma \xi ^{2}}-\frac{%
2+\xi }{\sigma \xi ^{2}}\left( 1+\frac{Y_{i}\xi }{\sigma }\right) ^{-1}+%
\frac{\left( 1+\xi \right) }{\sigma \xi ^{2}}\left( 1+\frac{Y_{i}\xi }{%
\sigma }\right) ^{-2}\right\} .
\end{eqnarray*}

Using the definition of GPD (\ref{GPD}), we have that%
\begin{eqnarray}
\mathbb{E}_{GPD}\left[ D_{i}|Y_{i}>u\right] &=&z^{-1/\xi }  \label{mean1} \\
\mathbb{E}_{GPD}\left[ \left. \left( 1+\frac{\xi Y_{i}}{\sigma }\right) ^{-r}%
\mathbf{1}\left[ Y_{i}\leq T\right] \right\vert Y_{i}>u\right] &=&\frac{%
1-z^{-r-1/\xi }}{1+r\xi }\text{ for any }r>0  \label{mean2} \\
\mathbb{E}_{GPD}\left[ \left. \log \left( 1+\frac{\xi Y_{i}}{\sigma }\right) 
\mathbf{1}\left[ Y_{i}\leq T\right] \right\vert Y_{i}>u\right] &=&\xi
-z^{-1/\xi }\left( \xi +\log z\right) .  \label{mean3}
\end{eqnarray}%
Then using (\ref{mean1})-(\ref{mean3}) to obtain that%
\begin{eqnarray*}
&&\mathbb{E}_{GPD}\left[ \left. -\frac{\partial ^{2}l_{i}\left( \xi ,\sigma
\right) }{\partial \xi ^{2}}\right\vert Y_{i}>u\right] \\
&=&z^{-1/\xi }\left\{ \frac{2}{\xi ^{3}}\log z-\frac{3}{\xi ^{3}}+\frac{4}{%
\xi ^{3}}z^{-1}-\frac{1}{\xi ^{3}}z^{-2}\right\} \\
&&+\frac{2}{\xi ^{3}}\left( \xi -z^{-1/\xi }\left( \xi +\log z\right)
\right) -\frac{3+\xi }{\xi ^{3}}(1-z^{-1/\xi }) \\
&&+\frac{2\left( 2+\xi \right) }{\xi ^{3}}\frac{1-z^{-1-1/\xi }}{1+\xi }-%
\frac{1+\xi }{\xi ^{3}}\frac{1-z^{-2-1/\xi }}{1+2\xi } \\
&=&\frac{2}{\left( 1+\xi \right) (1+2\xi )}+\frac{z^{-2-1/\xi }}{(1+\xi
)(1+2\xi )\xi ^{2}}\times \\
&&\left\{ -1-\xi +z(2+4\xi )-z^{2}(1+\xi )(1+2\xi )\right\} ,
\end{eqnarray*}%
\begin{eqnarray*}
&&\mathbb{E}_{GPD}\left[ \left. -\frac{\partial ^{2}l_{i}\left( \xi ,\sigma
\right) }{\partial \sigma ^{2}}\right\vert Y_{i}>u\right] \\
&=&z^{-1/\xi }\left\{ \frac{1}{\sigma ^{2}\xi }\left( 1-z^{-2}\right)
\right\} +\mathbb{E}_{GPD}\left[ (1-D_{i})\left\{ \frac{1}{\xi \sigma ^{2}}-%
\frac{1}{\sigma ^{2}}\left( 1+\frac{1}{\xi }\right) \left( 1+\frac{\xi Y_{i}%
}{\sigma }\right) ^{-2}\right\} \right] \\
&=&\frac{1}{\left( 1+2\xi \right) \sigma ^{2}}-\frac{z^{-2-1/\xi }}{\left(
1+2\xi \right) \sigma ^{2}},
\end{eqnarray*}%
and%
\begin{eqnarray*}
\mathbb{E}_{GPD}\left[ \left. -\frac{\partial ^{2}l_{i}\left( \xi ,\sigma
\right) }{\partial \sigma \partial \xi }\right\vert \right] &=&z^{-1/\xi
}\left\{ \frac{1}{\sigma \xi ^{2}}\left( 1-2z^{-1}+z^{-2}\right) \right\} \\
&&+\left( 1-z^{-1/\xi }\right) \frac{1}{\sigma \xi ^{2}}-\frac{2+\xi }{%
\sigma \xi ^{2}}\frac{1-z^{-1-1/\xi }}{1+\xi }+\frac{\left( 1+\xi \right) }{%
\sigma \xi ^{2}}\frac{1-z^{-2-1/\xi }}{1+2\xi } \\
&=&\frac{1}{\left( 1+\xi \right) (1+2\xi )\sigma }+\frac{z^{-2-1/\xi }}{%
\left( 1+\xi \right) (1+2\xi )\xi ^{2}\sigma }\times \\
&&\left\{ -\left( 1+\xi \right) ^{2}+\left( 1-2z\right) \left( 1+\xi \right)
(1+2\xi )+z\left( 2+\xi \right) \left( 1+2\xi \right) \right\} .
\end{eqnarray*}%
This completes the proof. $\blacksquare $

\begin{lemma}
\label{lemma SR2}Suppose Condition \ref{cond Hall82} holds. Then for any
positive-valued integrable function $h\left( \cdot \right) $ on $\left(
1,\infty \right) $ and any $c\in \left( 1,\infty \right) $, there exists
some function $\phi (u)\rightarrow 0$ as $u\rightarrow \infty $ such that 
\begin{equation*}
\int_{1}^{c}h\left( t\right) \frac{L\left( tu\right) }{L\left( u\right) }%
dt=\int_{1}^{c}h\left( t\right) dt+\phi \left( u\right) \int_{1}^{c}h\left(
t\right) k\left( t\right) dt+o\left( \phi \left( u\right) \right) .
\end{equation*}
\end{lemma}

\paragraph{Proof of Lemma \protect\ref{lemma SR2}}

The assumption on $L\left( \cdot \right) $ is sufficient for the SR2
assumption in \cite{Goldie87} and \cite{Smith87}. Therefore, we have%
\begin{equation*}
L\left( tu\right) /L\left( u\right) =1+k\left( t\right) \phi \left( u\right)
+o\left( \phi \left( u\right) \right) ,
\end{equation*}%
where $k\left( t\right) =C\int_{1}^{t}u^{-\beta -1}du$ for some constant $C$
and $\phi \left( u\right) =u^{-\beta }$ (pp.1179-1181 in \cite{Smith87}).
Then it suffices to show that 
\begin{equation*}
\frac{h\left( t\right) \{L\left( tu\right) /L\left( u\right) -1\}}{\phi
\left( u\right) }
\end{equation*}%
is uniformly dominated by some integrable function of $t$ as $u\rightarrow
\infty $. This is done by noting that%
\begin{eqnarray*}
\left\vert \frac{h\left( t\right) \{L\left( tu\right) /L\left( u\right) -1\}%
}{\phi \left( u\right) }\right\vert &\leq &C\left\vert h\left( t\right)
\left( 1-t^{-\beta }\right) \right\vert \\
&\leq &C\left\vert h\left( t\right) \right\vert
\end{eqnarray*}%
for $t>1$. $\blacksquare $

\paragraph*{Proof of Proposition \protect\ref{prop GPDindex}}

Denote $S_{k}\left( \xi ,\sigma \right) $ as the $2\times 1$ vector with
components $-\left( \partial /\partial \xi \right)
\dsum_{i=1}^{k}l_{i}\left( \xi ,\sigma \right) $ and $-\sigma \left(
\partial /\partial \sigma \right) \dsum_{i=1}^{k}l_{i}\left( \xi ,\sigma
\right) $, where%
\begin{equation*}
l_{i}(\xi ,\sigma )=D_{i}\log \left( 1-G\left( T_{u};\xi ,\sigma \right)
\right) +\left( 1-D_{i}\right) \log g\left( Y_{i};\xi ,\sigma \right)
\end{equation*}%
with $T_{u}=T-u$. Also denote%
\begin{equation*}
M_{k}\left( \xi ,\sigma \right) =k^{-1}\left[ 
\begin{array}{cc}
-\dsum_{i=1}^{k}\frac{\partial l_{i}^{2}\left( \xi ,\sigma \right) }{%
\partial \xi ^{2}} & -\sigma \dsum_{i=1}^{k}\frac{\partial l_{i}^{2}\left(
\xi ,\sigma \right) }{\partial \xi \partial \sigma } \\ 
-\sigma \dsum_{i=1}^{k}\frac{\partial l_{i}^{2}\left( \xi ,\sigma \right) }{%
\partial \xi \partial \sigma } & -\sigma ^{2}\dsum_{i=1}^{k}\frac{\partial
^{2}l_{i}\left( \xi ,\sigma \right) }{\partial \sigma ^{2}}%
\end{array}%
\right] .
\end{equation*}%
Then the intermediate value theorem yields that%
\begin{equation*}
k^{1/2}\binom{\hat{\xi}-\xi }{\hat{\sigma}/\sigma -1}=M_{k}\left( \dot{\xi},%
\dot{\sigma}\right) ^{-1}\left( k^{-1/2}S_{k}\left( \xi ,\sigma \right)
+o_{p}\left( 1\right) \right)
\end{equation*}%
for some intermediate values $\dot{\xi}$ and $\dot{\sigma}$. We next show
that (i) $k^{-1/2}S_{k}(\xi ,\sigma )$ converges to the normal random
variable by using Lyapunov Central Limit Theorem (CLT) and (ii) $M_{k}(%
\tilde{\xi},\tilde{\sigma})$ uniformly converges to $M$ over $(\tilde{\xi},%
\tilde{\sigma})$ in a shrinking neighborhood centered at $(\xi ,\sigma )$.

To show (i), we use Lemma \ref{lemma SR2} and integration by parts to obtain
that, for any $r>0$ 
\begin{eqnarray}
&&\mathbb{E}\left[ \left. \left( 1+\frac{Y_{i}}{u}\right) ^{-r}\mathbf{1}%
\left[ Y_{i}\leq T\right] \right\vert Y_{i}>u\right]  \label{EY} \\
&=&-\int_{0}^{T-u}\left( 1+\frac{y}{u}\right) ^{-r}d\left( 1-F_{u}\left(
y\right) \right)  \notag \\
&=&\left. -\left( 1+\frac{y}{u}\right) ^{-r}\left( 1-F_{u}\left( y\right)
\right) \right\vert _{0}^{T-u}-r\int_{0}^{T-u}\left( 1-F_{u}\left( y\right)
\right) \left( 1+\frac{y}{u}\right) ^{-r-1}u^{-1}dy  \notag \\
&=&1-\left( 1+\frac{T-u}{u}\right) ^{-r}\left( 1-F_{u}\left( T-u\right)
\right)  \notag \\
&&-r\int_{1}^{\frac{T-u}{u}+1}t^{-r-1-\alpha }\frac{L\left( ut\right) }{%
L\left( u\right) }dt  \notag \\
&=&1-(T/u)^{-r-\alpha }+\frac{r}{r+\alpha }\left( (T/u)^{-r-\alpha
}-1\right) +O\left( \phi \left( u\right) \right)  \notag \\
&=&\frac{\alpha }{r+\alpha }\left( 1-(T/u)^{-r-\alpha }\right) +O\left( \phi
\left( u\right) \right)  \notag
\end{eqnarray}%
and%
\begin{eqnarray}
&&\mathbb{E}\left[ \left. \log \left( 1+\frac{Y_{i}}{u}\right) \mathbf{1}%
\left[ Y_{i}\leq T\right] \right\vert Y_{i}>u\right]  \label{ElogY} \\
&=&-\int_{0}^{T-u}\log \left( 1+\frac{y}{u}\right) d\left( 1-F_{u}\left(
y\right) \right)  \notag \\
&=&\left. -\log \left( 1+\frac{y}{u}\right) \left( 1-F_{u}\left( y\right)
\right) \right\vert _{0}^{T-u}+\int_{0}^{T-u}\left( 1-F_{u}\left( y\right)
\right) \left( 1+\frac{y}{u}\right) ^{-1}u^{-1}dy  \notag \\
&=&-\log \left( 1+\frac{T-u}{u}\right) \left( 1-F_{u}\left( T-u\right)
\right) +\int_{1}^{\frac{T-u}{u}+1}\frac{L\left( ut\right) }{L\left(
u\right) }t^{-1-\alpha }dt  \notag \\
&=&-\left( \log (T/u)\right) (T/u)^{-\alpha }+\frac{1}{\alpha }\left(
1-(T/u)^{-\alpha }\right) +O\left( \phi \left( u\right) \right) ,  \notag
\end{eqnarray}%
where recall $\phi \left( u\right) =u^{-\beta }$. Use the change of variable 
$\sigma =u/\alpha $ and recall $\alpha =1/\xi $. Then by Lemma \ref{lemma
GPD}, we have $z=1+\frac{T-u}{\sigma }\xi =T/u$ and%
\begin{eqnarray*}
&&\mathbb{E}\left[ \left. -\frac{\partial l_{i}\left( \xi ,\sigma \right) }{%
\partial \xi }\right\vert Y_{i}>u\right] \\
&=&\mathbb{E}\left[ D_{i}|Y_{i}>u\right] \left\{ -\frac{1}{\xi ^{2}}\log z+%
\frac{1}{\xi ^{2}}\left( 1-z^{-1}\right) \right\} \\
&&+\mathbb{E}\left[ \left. \left\{ -\frac{1}{\xi ^{2}}\log \left( 1+\frac{%
\xi }{\sigma }Y_{i}\right) +\frac{1}{\xi }\left( 1+\frac{1}{\xi }\right)
\left( 1-\left( 1+\frac{\xi }{\sigma }Y_{i}\right) ^{-1}\right) \right\} 
\mathbf{1}\left[ Y_{i}\leq T\right] \right\vert Y_{i}>u\right] \\
&=&\left( 1-F_{u}\left( T-u\right) \right) \left\{ -\frac{1}{\xi ^{2}}\log
(T/u)+\frac{1}{\xi ^{2}}\left( 1-(T/u)^{-1}\right) \right\} \\
&&-\frac{1}{\xi ^{2}}\mathbb{E}\left[ \left. \log \left( 1+\frac{Y_{i}}{u}%
\right) \mathbf{1}\left[ Y_{i}\leq T\right] \right\vert Y_{i}>u\right] \\
&&+\mathbb{E}\left[ \left. \frac{1}{\xi }\left( 1+\frac{1}{\xi }\right)
\left( 1-\left( 1+\frac{Y_{i}}{u}\right) ^{-1}\right) \mathbf{1}\left[
Y_{i}\leq T\right] \right\vert Y_{i}>u\right] \\
&=&(T/u)^{-\alpha }\left\{ -\alpha ^{2}\log (T/u)+\alpha ^{2}\left(
1-(T/u)^{-1}\right) \right\} -\alpha ^{2}\left( -\left( \log (T/u)\right)
(T/u)^{-\alpha }+\frac{1}{\alpha }\left( 1-(T/u)^{-\alpha }\right) \right) \\
&&+\alpha \left( 1+\alpha \right) \left( 1-(T/u)^{-\alpha }\right) -\alpha
^{2}\left( 1-(T/u)^{-1-\alpha }\right) +O\left( \phi \left( u\right) \right)
\\
&=&O\left( \phi \left( u\right) \right) .
\end{eqnarray*}%
Similarly by Condition \ref{cond top} and repetitively using (\ref{EY}) and (%
\ref{ElogY}), we have that%
\begin{eqnarray*}
\mathbb{E}\left[ \left. -\sigma \frac{\partial l_{i}\left( \xi ,\sigma
\right) }{\partial \sigma }\right\vert Y_{i}>u\right] &=&O\left( \phi \left(
u\right) \right) \\
\mathbb{E}\left[ \left. \left( \frac{\partial l_{i}\left( \xi ,\sigma
\right) }{\partial \xi }\right) ^{2}\right\vert Y_{i}>u\right]
&=&M_{11}+O\left( \phi \left( u\right) \right) +o(1) \\
\mathbb{E}\left[ \left. \sigma ^{2}\left( \frac{\partial l_{i}\left( \xi
,\sigma \right) }{\partial \sigma ^{2}}\right) ^{2}\right\vert Y_{i}>u\right]
&=&M_{22}+O\left( \phi \left( u\right) \right) +o(1) \\
\mathbb{E}\left[ \left. \sigma \frac{\partial l_{i}\left( \xi ,\sigma
\right) }{\partial \sigma }\frac{\partial l_{i}\left( \xi ,\sigma \right) }{%
\partial \xi }\right\vert Y_{i}>u\right] &=&M_{12}+O\left( \phi \left(
u\right) \right) +o(1),
\end{eqnarray*}%
and the third moments conditional on $Y_{i}>u$ of $\partial l_{i}\left( \xi
,\sigma \right) /\partial \xi $ and $\sigma \partial l_{i}\left( \xi ,\sigma
\right) /\partial \sigma $ are also bounded as $u\rightarrow \infty $. Then
by Lyapunov CLT, we have $k^{-1/2}S_{k}\left( \xi ,\sigma \right) \overset{d}%
{\rightarrow }\mathcal{N}\left( 0,M\right) $.

Now it remains to show that $M_{k}(\tilde{\xi},\tilde{\sigma})$ uniformly
converges to $M$ in the neighborhood that $\{(\tilde{\sigma},\tilde{\xi}):$ $%
\left\vert \tilde{\sigma}/\sigma -1\right\vert \leq \varepsilon _{k}$ and $|%
\tilde{\xi}-\xi |\leq \varepsilon _{k}\}$ for some $\varepsilon _{k}=o(1)$
satisfying $k^{1/2}\varepsilon _{k}\rightarrow \infty $. To this end, it
suffices to show that $\mathbb{E[}|\partial ^{3}l_{i}(\tilde{\xi},\tilde{%
\sigma})/\partial \xi ^{3}||Y_{i}>u]$, $\mathbb{E[}|\sigma \partial
^{3}l_{i}(\tilde{\xi},\tilde{\sigma})/\partial \xi ^{2}\partial \sigma
||Y_{i}>u]$, $\mathbb{E[}|\sigma ^{3}\partial ^{3}l_{i}(\tilde{\xi},\tilde{%
\sigma})/\partial \sigma ^{3}||Y_{i}>u]$, and $\mathbb{E[}\sigma
^{2}|\partial ^{3}l_{i}(\tilde{\xi},\tilde{\sigma})/\partial \sigma
^{2}\partial \xi ||Y_{i}>u]$ are all uniformly bounded over this
neighborhood. This is done by straightforward calculations as we show in
Lemma \ref{lemma boundd3}. For brevity, we present the proof for $\mathbb{E[}%
|\partial ^{3}l_{i}(\tilde{\xi},\tilde{\sigma})/\partial \xi ^{3}||Y_{i}>u]$
only since the argument for the other terms are similar (cf.\ pp.1178-1180
in \cite{Smith87}). $\blacksquare $

\begin{lemma}
\label{lemma boundd3}$\mathbb{E[}|\partial ^{3}l_{i}(\tilde{\xi},\tilde{%
\sigma})/\partial \xi ^{3}||Y_{i}>u]$ is uniformly bounded over $\{(\tilde{%
\sigma},\tilde{\xi}):$ $\left\vert \tilde{\sigma}/\sigma -1\right\vert \leq
\varepsilon _{k}$ and $|\tilde{\xi}-\xi |\leq \varepsilon _{k}\}$ for any $%
\varepsilon _{k}\rightarrow 0$ as $k\rightarrow \infty $.
\end{lemma}

\paragraph{Proof of Lemma \protect\ref{lemma boundd3}}

Substituting (\ref{loglikeli}) to obtain that%
\begin{eqnarray*}
&&\frac{\partial ^{3}l_{i}\left( \tilde{\xi},\tilde{\sigma}\right) }{%
\partial \xi ^{3}} \\
&=&D_{i}\left\{ 
\begin{array}{c}
\frac{2T_{u}^{3}}{\tilde{\sigma}^{3}\tilde{\xi}}\left( 1+\frac{T_{u}\tilde{%
\xi}}{\tilde{\sigma}}\right) ^{-3}+\frac{3T_{u}^{2}}{\tilde{\sigma}^{2}%
\tilde{\xi}^{2}}\left( 1+\frac{T_{u}\tilde{\xi}}{\tilde{\sigma}}\right) ^{-2}
\\ 
+\frac{6T_{u}}{\tilde{\sigma}\tilde{\xi}^{3}}\left( 1+\frac{T_{u}\tilde{\xi}%
}{\tilde{\sigma}}\right) ^{-1}-\frac{6}{\tilde{\xi}^{4}}\log \left( 1+\frac{%
T_{u}\tilde{\xi}}{\tilde{\sigma}}\right)%
\end{array}%
\right\} \\
&&+(1-D_{i})\left\{ 
\begin{array}{l}
\frac{11}{\tilde{\xi}^{4}}+\frac{2}{\tilde{\xi}^{3}}-\frac{2}{\tilde{\xi}^{4}%
}\left( 1+\frac{\tilde{\xi}}{\tilde{\sigma}}Y_{i}\right) ^{-3}+\left( \frac{9%
}{\tilde{\xi}^{4}}+\frac{6}{\tilde{\xi}^{3}}\right) \left( 1+\frac{\tilde{\xi%
}}{\tilde{\sigma}}Y_{i}\right) ^{-2} \\ 
-\left( \frac{18}{\tilde{\xi}^{4}}+\frac{6}{\tilde{\xi}^{3}}\right) \left( 1+%
\frac{\tilde{\xi}}{\tilde{\sigma}}Y_{i}\right) ^{-1}-\frac{6}{\tilde{\xi}^{4}%
}\log \left( 1+\frac{\tilde{\xi}Y_{i}}{\tilde{\sigma}}\right)%
\end{array}%
\right\} \\
&\equiv &B_{1n}(\tilde{\xi},\tilde{\sigma})+B_{2n}(\tilde{\xi},\tilde{\sigma}%
).
\end{eqnarray*}

Since $T_{u}=T-u=O(\sigma )$ and $\mathbb{E}\left[ D_{i}|Y_{i}>u\right] <1$,
the expectation of $B_{1n}$ is then uniformly bounded over $\{(\tilde{\sigma}%
,\tilde{\xi}):$ $\left\vert \tilde{\sigma}/\sigma -1\right\vert \leq
\varepsilon _{k}$ and $|\tilde{\xi}-\xi |\leq \varepsilon _{k}\}$. To bound $%
B_{2n}$, it suffices to show that for any $\tilde{\xi}$ and $\tilde{\sigma}$
in the set $\{(\tilde{\sigma},\tilde{\xi}):$ $\left\vert \tilde{\sigma}%
/\sigma -1\right\vert \leq \varepsilon _{k}$ and $|\tilde{\xi}-\xi |\leq
\varepsilon _{k}\}$ and for any $r>0$, 
\begin{eqnarray}
&&\mathbb{E}\left[ \left. \left( 1+\frac{\tilde{\xi}}{\tilde{\sigma}}%
Y_{i}\right) ^{-r}\mathbf{1}\left[ Y_{i}\leq T\right] \right\vert Y_{i}>u%
\right] \left. <\right. \infty \text{ and}  \label{EYt} \\
&&\mathbb{E}\left[ \left. \log \left( 1+\frac{\tilde{\xi}Y_{i}}{\tilde{\sigma%
}}\right) \mathbf{1}\left[ Y_{i}\leq T\right] \right\vert Y_{i}>u\right]
\left. <\right. \infty ,  \label{ElogYt}
\end{eqnarray}%
which are similar to (\ref{EY}) and (\ref{ElogY}), respectively. In
particular, denote $v=\left( \xi \tilde{\sigma}\right) /(\tilde{\xi}\sigma )$
so that $|v-1|\leq \varepsilon $ for some constant $\varepsilon \rightarrow
0 $. Then using the change of variable $\sigma =u/\alpha =u\xi $, we have 
\begin{eqnarray*}
&&\mathbb{E}\left[ \left. \left( 1+\frac{\tilde{\xi}}{\tilde{\sigma}}%
Y_{i}\right) ^{-r}\mathbf{1}\left[ Y_{i}\leq T\right] \right\vert Y_{i}>u%
\right] \\
&=&\mathbb{E}\left[ \left. \left( 1+\frac{Y_{i}}{vu}\right) ^{-r}\mathbf{1}%
\left[ Y_{i}\leq T\right] \right\vert Y_{i}>u\right] \\
&=&-\int_{0}^{T-u}\left( 1+\frac{y}{vu}\right) ^{-r}d\left( 1-F_{u}\left(
y\right) \right) \\
&=&\left. -\left( 1+\frac{y}{vu}\right) ^{-r}\left( 1-F_{u}\left( y\right)
\right) \right\vert _{0}^{T-u}-vr\int_{0}^{T-u}\left( 1-F_{u}\left( y\right)
\right) \left( 1+\frac{y}{vu}\right) ^{-r-1}u^{-1}dy \\
&=&1-\left( 1+\frac{T-u}{vu}\right) ^{-r}\left( 1-F_{u}\left( T-u\right)
\right) \\
&&-vr\int_{1}^{\frac{T-u}{u}+1}\left( 1+\frac{t-1}{v}\right)
^{-r-1}t^{-\alpha }\frac{L\left( ut\right) }{L\left( u\right) }dt.
\end{eqnarray*}%
Condition \ref{cond top} and Lemma \ref{lemma SR2} yield that the above item
is bounded. A very similar argument applies to (\ref{ElogYt}), which
completes the proof. $\blacksquare $

\paragraph{Proof of Proposition \protect\ref{prop GPDquan}}

The proof follows analogously from Theorem 4.3.1 and Remark 4.3.7 in \cite%
{deHaan07}. We now provide the details. Recall $d_{n}=\left( m+k\right)
/(np_{n})$ and write $u=\hat{Q}\left( 1-p_{n}d_{n}\right) $, which is $%
Y_{(m+k)}$. Then we decompose $\hat{Q}\left( 1-p_{n}\right) -Q(1-p_{n})$ as 
\begin{equation*}
\sqrt{k}\frac{\hat{Q}\left( 1-p_{n}\right) -Q(1-p_{n})}{\sigma q_{\xi
}(d_{n})}=C_{1n}+C_{2n}+C_{3n}-C_{4n},
\end{equation*}%
where%
\begin{eqnarray*}
C_{1n} &=&\sqrt{k}\frac{\hat{Q}\left( 1-p_{n}d_{n}\right) -Q(1-p_{n}d_{n})}{%
\sigma }\frac{1}{q_{\xi }(d_{n})} \\
C_{2n} &=&\frac{\hat{\sigma}}{\sigma }\left\{ \frac{\sqrt{k}}{q_{\xi }(d_{n})%
}\left( \frac{d_{n}^{-\hat{\xi}}-1}{\hat{\xi}}-\frac{d_{n}^{-\xi }-1}{\xi }%
\right) \right\} \\
C_{3n} &=&\sqrt{k}\left( \frac{\hat{\sigma}}{\sigma }-1\right) \frac{%
d_{n}^{-\xi }-1}{\xi q_{\xi }(d_{n})} \\
C_{4n} &=&\frac{\sqrt{k}}{q_{\xi }(d_{n})}\left( \frac{%
Q(1-p_{n})-Q(1-p_{n}d_{n})}{\sigma }-\frac{d_{n}^{-\xi }-1}{\xi }\right) .
\end{eqnarray*}%
We next derive the limits of $C_{jn}$ for $j=1,2,3,4$. To this end, we
define $U\left( t\right) =Q(1-1/t)$ and denote $U^{\prime }(t)=\partial
U(t)/\partial t$. We introduce the second-order tail approximation that%
\begin{equation}
\lim_{t\rightarrow \infty }\frac{\frac{U(ty)-U(t)}{a(t)}-\frac{y^{\xi }-1}{%
\xi }}{A(t)}=H(y)  \label{2nd cond}
\end{equation}%
as in Theorem 2.3.12 in \cite{deHaan07}. Condition \ref{cond Hall82} implies
that $a(t)=tU^{\prime }(t)=1/\left( tf(Q(1-1/t))\right) $, $A(t)\propto
t^{-\beta /\alpha }$, and $H(y)=-y^{\xi }(y^{-\beta }-1)/\beta $.

For $C_{1n}$, substitute $t=1/\left( p_{n}d_{n}\right) $ and use Theorem
2.4.1 in \cite{deHaan07} to obtain that%
\begin{equation*}
\sqrt{k}\frac{\left( \hat{Q}\left( 1-p_{n}d_{n}\right)
-Q(1-p_{n}d_{n})\right) }{a_{n}(1/(p_{n}d_{n}))}\overset{d}{\rightarrow }%
\mathcal{N}\left( 0,1\right) .
\end{equation*}%
Then by Theorem 1.1.6 in \cite{deHaan07}, we have that $\sigma $ is
asymptotically equivalent to $a(1/(p_{n}d_{n}))$ as $n\rightarrow \infty $,
which further implies that $C_{1n}\overset{d}{\rightarrow }\mathcal{N}\left(
0,q_{\xi }(d_{0})^{-2}\right) $.

For $C_{2n}$, the same argument as part II on pp.136-137 in \cite{deHaan07}
yields that $C_{2n}=k^{1/2}\left( \hat{\xi}-\xi \right) +o_{p}(1)$. For $%
C_{3n}$, by Proposition \ref{prop GPDindex}, we have that 
\begin{equation*}
C_{3n}=k^{1/2}\left( \frac{\hat{\sigma}}{\sigma }-1\right) \left( \frac{%
d_{0}^{-\xi }-1}{\xi q_{\xi }(d_{0})}\right) +o_{p}(1),
\end{equation*}%
where recall $d_{0}=\lim_{n\rightarrow \infty }d_{n}>0$. Note that given $%
u=Y_{(m+k)}$, the excesses $\{Y_{(m+i)}-Y_{(m+k)}\}_{i=1}^{k-1}$ are
asymptotically independent from $Y_{(m+k)}$ (cf. p.1185 in \cite{Drees04}),
and therefore $C_{1n}$ is asymptotically independent from $C_{2n}$ and $%
C_{3n}$ (see also pp.1180-1181 in \cite{Smith87}).

Finally, (\ref{2nd cond}) and Condition \ref{cond kn} yield that $\sqrt{k}%
A(n/(k+m))=o(1)$ and hence 
\begin{eqnarray*}
C_{4n} &=&\sqrt{k}A(n/(k+m))\frac{d_{n}^{-\xi }-1}{\xi q_{\xi }(d_{n})}\frac{%
\left( \frac{U(1/p_{n})-U(1/\left( p_{n}d_{n}\right) )}{a(n/(m+k))}\frac{\xi 
}{d_{n}^{-\xi }-1}-1\right) }{A(n/(k+m))} \\
&=&o(1).
\end{eqnarray*}%
The proof is complete by combining $C_{jn}$ for $j=1,2,3,4$. $\blacksquare $

\paragraph{Proof of Proposition \protect\ref{prop evt}}

We prove this by induction. By standard EVT, for any fixed positive integer $%
I$,%
\begin{equation}
f_{X_{1},\ldots ,X_{I}|\xi }(x_{1},\ldots ,x_{I})=V_{\xi }\left(
x_{I}\right) \tprod_{i=1}^{I}v_{\xi }\left( x_{i}\right) /V_{\xi }\left(
x_{i}\right) .  \label{evt}
\end{equation}%
Consider $m=1$ first. For any fixed positive integer $k$, (\ref{evt}) with $%
I=k+1$ implies that 
\begin{eqnarray*}
f_{X_{m+1},\ldots ,X_{m+k}|\xi }(x_{m+1},\ldots ,x_{m+k})
&=&f_{X_{2},...,X_{k+1}|\xi }(x_{2},\ldots ,x_{k+1}) \\
&=&\left( \int_{x_{2}}^{\infty }\frac{v_{\xi }\left( x_{1}\right) }{V_{\xi
}\left( x_{1}\right) }dx_{1}\right) V_{\xi }\left( x_{k+1}\right)
\tprod_{i=2}^{k+1}v_{\xi }\left( x_{i}\right) /V_{\xi }\left( x_{i}\right) \\
&=&-\log V_{\xi }\left( x_{2}\right) V_{\xi }\left( x_{k+1}\right)
\tprod_{i=2}^{k+1}v_{\xi }\left( x_{i}\right) /V_{\xi }\left( x_{i}\right) ,
\end{eqnarray*}%
which satisfies (\ref{fx}).

Now assume (\ref{fx}) holds for some fixed positive integer $m\geq 1$. This
implies that for any $k$,%
\begin{eqnarray*}
&&f_{X_{m+2},\ldots ,X_{m+1+k}|\xi }(x_{m+2},\ldots ,x_{m+k+1}) \\
&=&\int_{x_{m+2}}^{\infty }f_{X_{m+1},...,X_{m+k+1}}(x_{m+1},\ldots
,x_{m+k+1})dx_{m+1} \\
&=&\left( \int_{x_{m+2}}^{\infty }\frac{1}{m!}\left( -\log V_{\xi }\left(
x_{m+1}\right) \right) ^{m}\frac{v_{\xi }(x_{m+1})}{V_{\xi }(x_{m+1})}%
dx_{m+1}\right) V_{\xi }\left( x_{m+k+1}\right) \tprod_{i=m+2}^{m+k+1}v_{\xi
}\left( x_{i}\right) /V_{\xi }\left( x_{i}\right) \\
&=&\left( \int_{\log G_{\xi }\left( x_{m+2}\right) }^{0}\left( -v\right)
^{m}dv\right) \frac{1}{m!}V_{\xi }\left( x_{m+k+1}\right)
\tprod_{i=m+2}^{m+k+1}v_{\xi }\left( x_{i}\right) /V_{\xi }\left(
x_{i}\right) \\
&=&\frac{1}{\left( m+1\right) !}\left( -\log V_{\xi }\left( x_{m+2}\right)
\right) ^{m+1}V_{\xi }\left( x_{m+k+1}\right) \tprod_{i=m+2}^{m+k+1}v_{\xi
}\left( x_{i}\right) /V_{\xi }\left( x_{i}\right) ,
\end{eqnarray*}%
which means that (\ref{fx}) holds for $m+1$. This completes the proof. $%
\blacksquare $

\subsection*{A.3 Additional Empirical Results in CPS Data}

Tables \ref{tbl cps 008} and \ref{tbl cps 010} depict the results based on $%
k=\left[ 0.04n\right] $ and $\left[ 0.06n\right] $, respectively.

\begin{table}[H]
\begin{center}%
\caption{Empirical Results Using 2019 March CPS Data}\label{tbl cps 008}%
\vspace{+2ex}%

\begin{tabular}{cccccccccc}
\hline\hline
\multicolumn{10}{c}{Panel A: 95\% confidence intervals with race-based
subsample} \\ 
& $n$ & cen\# & cen\% & \multicolumn{2}{c}{tail index} & \multicolumn{2}{c}{
Q(0.99)} & \multicolumn{2}{c}{Q(0.999)} \\ \hline
\multicolumn{1}{l}{Full Sample} & \multicolumn{1}{r}{115424} & 672 & 0.58 & 
(0.37 & 0.49) & (24.26 & 25.32) & (59.88 & 73.00) \\ 
\multicolumn{1}{l}{Male} & \multicolumn{1}{r}{55553} & 491 & 0.88 & (0.39 & 
0.62) & (28.76 & 30.97) & (71.86 & 106.3) \\ 
\multicolumn{1}{l}{Female} & \multicolumn{1}{r}{59871} & 181 & 0.30 & (0.30
& 0.44) & (18.31 & 19.33) & (42.46 & 52.03) \\ 
\multicolumn{1}{l}{Male White} & \multicolumn{1}{r}{43371} & 419 & 0.97 & 
(0.09 & 0.34) & (29.57 & 31.84) & (54.54 & 75.61) \\ 
\multicolumn{1}{l}{Female White} & \multicolumn{1}{r}{45424} & 141 & 0.31 & 
(0.29 & 0.45) & (18.40 & 19.59) & (42.36 & 53.68) \\ 
\multicolumn{1}{l}{Male Asian} & \multicolumn{1}{r}{3676} & 50 & 1.36 & (0.00
& 0.55) & (30.73 & 37.77) & (48.30 & 109.6) \\ 
\multicolumn{1}{l}{Female Asian} & \multicolumn{1}{r}{4099} & 22 & 0.54 & 
(0.13 & 0.76) & (21.27 & 26.66) & (42.89 & 104.5) \\ 
\multicolumn{1}{l}{Male Hispanic} & \multicolumn{1}{r}{44420} & 445 & 1.00 & 
(0.11 & 0.36) & (29.98 & 32.27) & (55.56 & 77.78) \\ 
\multicolumn{1}{l}{Female Hispanic} & \multicolumn{1}{r}{48192} & 155 & 0.32
& (0.38 & 0.55) & (18.83 & 19.99) & (45.77 & 59.36) \\ 
\multicolumn{1}{l}{Male Black} & \multicolumn{1}{r}{6144} & 12 & 0.20 & (0.16
& 0.53) & (16.06 & 18.55) & (29.70 & 47.84) \\ 
\multicolumn{1}{l}{Female Black} & \multicolumn{1}{r}{7827} & 9 & 0.16 & 
(0.12 & 0.44) & (13.94 & 15.72) & (25.08 & 36.81) \\ \hline
\multicolumn{10}{c}{Panel B: 95\% confidence intervals with age-based
subsample} \\ 
\multicolumn{1}{l}{Age} & $n$ & cen\# & cen\% & \multicolumn{2}{c}{tail index
} & \multicolumn{2}{c}{Q(0.99)} & \multicolumn{2}{c}{Q(0.999)} \\ \hline
\multicolumn{1}{l}{18-30} & 27829 & 35 & 0.13 & (0.34 & 0.53) & (12.63 & 
13.54) & (28.42 & 37.19) \\ 
\multicolumn{1}{l}{30-40} & 25213 & 158 & 0.63 & (0.24 & 0.51) & (24.15 & 
26.40) & (50.57 & 75.56) \\ 
\multicolumn{1}{l}{40-50} & 23419 & 213 & 0.91 & (0.00 & 0.32) & (28.67 & 
31.48) & (48.07 & 71.41) \\ 
\multicolumn{1}{l}{50-60} & 21767 & 196 & 0.90 & (0.63 & 1.00) & (28.30 & 
32.89) & (86.35 & 218.9) \\ 
\multicolumn{1}{l}{60-65} & 17196 & 70 & 0.41 & (0.43 & 0.76) & (20.29 & 
22.63) & (49.71 & 88.34) \\ \hline
\end{tabular}

\vspace{-4ex}%
\end{center}
\begin{singlespacing}%
Note: Entries are the sample size ($n$), the number of censored observations
(cen\#), the censored fraction in percentage points (cen\%),\ 95\%
confidence intervals of the tail index and those of the 99\% and 99.9\%
quantiles measured in 10$^{4}$\ USD. The results are based on $k=\left[ 0.04n%
\right] $ and the variable ERN\_VAL in the CPS dataset (and equivalently the
variable inclongj from the IPUMS dataset). Data are available at
https://cps.ipums.org/cps. 
\end{singlespacing}%
\end{table}%

\begin{table}[H]
\begin{center}%
\caption{Empirical Results Using 2019 March CPS Data}\label{tbl cps 010}%
\vspace{+2ex}%

\begin{tabular}{cccccccccc}
\hline\hline
\multicolumn{10}{c}{Panel A: 95\% confidence intervals in race-based
subsamples} \\ 
& $n$ & cen\# & cen\% & \multicolumn{2}{c}{tail index} & \multicolumn{2}{c}{
Q(0.99)} & \multicolumn{2}{c}{Q(0.999)} \\ \hline
\multicolumn{1}{l}{Full Sample} & \multicolumn{1}{r}{115424} & 672 & 0.58 & 
(0.31 & 0.40) & (24.33 & 25.34) & (56.01 & 65.09) \\ 
\multicolumn{1}{l}{Male} & \multicolumn{1}{r}{55553} & 491 & 0.88 & (0.30 & 
0.45) & (28.59 & 30.52) & (63.88 & 82.90) \\ 
\multicolumn{1}{l}{Female} & \multicolumn{1}{r}{59871} & 181 & 0.30 & (0.43
& 0.54) & (18.03 & 19.05) & (46.57 & 57.68) \\ 
\multicolumn{1}{l}{Male White} & \multicolumn{1}{r}{43371} & 419 & 0.97 & 
(0.32 & 0.50) & (29.55 & 31.96) & (67.36 & 93.12) \\ 
\multicolumn{1}{l}{Female White} & \multicolumn{1}{r}{45424} & 141 & 0.31 & 
(0.43 & 0.57) & (18.09 & 19.30) & (47.07 & 60.62) \\ 
\multicolumn{1}{l}{Male Asian} & \multicolumn{1}{r}{3676} & 50 & 1.36 & (0.00
& 0.51) & (30.59 & 38.09) & (48.25 & 102.7) \\ 
\multicolumn{1}{l}{Female Asian} & \multicolumn{1}{r}{4099} & 22 & 0.54 & 
(0.16 & 0.61) & (21.27 & 26.54) & (41.85 & 87.70) \\ 
\multicolumn{1}{l}{Male Hispanic} & \multicolumn{1}{r}{44420} & 445 & 1.00 & 
(0.43 & 0.61) & (30.25 & 32.94) & (78.16 & 113.6) \\ 
\multicolumn{1}{l}{Female Hispanic} & \multicolumn{1}{r}{48192} & 155 & 0.32
& (0.38 & 0.51) & (18.85 & 20.01) & (45.79 & 57.46) \\ 
\multicolumn{1}{l}{Male Black} & \multicolumn{1}{r}{6144} & 12 & 0.20 & (0.19
& 0.50) & (16.05 & 18.31) & (28.61 & 46.11) \\ 
\multicolumn{1}{l}{Female Black} & \multicolumn{1}{r}{7827} & 9 & 0.16 & 
(0.20 & 0.47) & (13.83 & 15.56) & (25.73 & 38.47) \\ \hline
\multicolumn{10}{c}{Panel B: 95\% confidence intervals in age-based
subsamples} \\ 
\multicolumn{1}{l}{Age} & $n$ & cen\# & cen\% & \multicolumn{2}{c}{tail index
} & \multicolumn{2}{c}{Q(0.99)} & \multicolumn{2}{c}{Q(0.999)} \\ \hline
\multicolumn{1}{l}{18-30} & 27829 & 35 & 0.13 & (0.30 & 0.44) & (12.78 & 
13.68) & (27.65 & 34.78) \\ 
\multicolumn{1}{l}{30-40} & 25213 & 158 & 0.63 & (0.57 & 0.79) & (23.95 & 
26.70) & (71.84 & 118.8) \\ 
\multicolumn{1}{l}{40-50} & 23419 & 213 & 0.91 & (0.27 & 0.50) & (28.59 & 
31.63) & (59.61 & 90.30) \\ 
\multicolumn{1}{l}{50-60} & 21767 & 196 & 0.90 & (0.35 & 0.59) & (28.12 & 
31.56) & (65.11 & 105.7) \\ 
\multicolumn{1}{l}{60-65} & 17196 & 70 & 0.41 & (0.40 & 0.64) & (20.54 & 
22.99) & (50.78 & 80.72) \\ \hline
\end{tabular}

\vspace{-4ex}%
\end{center}
\begin{singlespacing}%
Note: Entries are the sample size ($n$), the number of censored observations
(cen\#), the censored fraction in percentage points (cen\%),\ 95\%
confidence intervals of the tail index and those of the 99\% and 99.9\%
quantiles measured in 10$^{4}$\ USD. The results are based on $k=\left[ 0.06n%
\right] $ and the variable ERN\_VAL in the CPS dataset (and equivalently the
variable inclongj from the IPUMS dataset). Data are available at
https://cps.ipums.org/cps. 
\end{singlespacing}%
\end{table}%

\newpage

\bibliographystyle{econometrica}
\bibliography{diss}

\end{document}